\documentclass[10pt, journal, compsoc]{IEEEtran}
\usepackage{amsmath,amsfonts}
\usepackage{array}
\usepackage{textcomp}
\usepackage{stfloats}
\usepackage{url}
\usepackage{verbatim}
\usepackage{graphicx}
\def\BibTeX{{\rm B\kern-.05em{\sc i\kern-.025em b}\kern-.08em
    T\kern-.1667em\lower.7ex\hbox{E}\kern-.125emX}}

\usepackage{
    graphics,
    booktabs,
    breakurl,
    subcaption,
    multirow,
    float,
    bm,
    balance,
    amssymb,
    mathtools,
    xcolor,
    array,
    algorithm,
    algpseudocode,
    pifont,
    threeparttable,
    cite
}

\begin{document}

\newcommand{\tdsc}[1]{{\color{blue}{#1}}}

\title{


Phoneme-Based Proactive Anti-Eavesdropping with Controlled Recording Privilege
}

\author{
Peng Huang, Yao Wei, Peng Cheng, Zhongjie Ba,~\IEEEmembership{Member,~IEEE}, Li Lu, Feng Lin,~\IEEEmembership{Member,~IEEE}, Yang Wang, and Kui Ren~\IEEEmembership{Fellow,~IEEE}

\IEEEcompsocitemizethanks{
\IEEEcompsocthanksitem 
Peng Huang, Yao Wei, Peng Cheng, Zhongjie Ba, Li Lu, Feng Lin, Yang Wang, and Kui Ren
are with the State Key Laboratory of Blockchain and Data Security, Zhejiang University, Hangzhou, Zhejiang 310000, China.
E-mail: {penghuang, weiy, peng\_cheng, zhongjieba, li.lu, flin, Kjchz, kuiren}@zju.edu.cn.
}
\thanks{Manuscript received April XX, 20XX; revised August XX, 20XX.

(Corresponding author: Zhongjie Ba.)
}
}

\markboth{Journal of \LaTeX\ Class Files,~Vol.~14, No.~8, August~20xx}%
{Shell \MakeLowercase{\textit{et al.}}: A Sample Article Using IEEEtran.cls for IEEE Journals}

\maketitle

\begin{abstract}
    The widespread smart devices raise people's concerns of being eavesdropped on.
To enhance voice privacy, recent studies exploit the nonlinearity in microphone to jam audio recorders with inaudible ultrasound.
However, existing solutions solely rely on energetic masking. Their simple-form noise leads to several problems, such as high energy requirements and being easily removed by speech enhancement techniques. Besides, most of these solutions do not support authorized recording, which restricts their usage scenarios.
In this paper, we design an efficient yet robust system that can jam microphones while preserving authorized recording.
Specifically, we propose a novel phoneme-based noise with the idea of informational masking, which can distract both machines and humans and is resistant to denoising techniques. 
Besides, we optimize the noise transmission strategy for broader coverage and implement a hardware prototype of our system.
Experimental results show that our system can reduce the recognition accuracy of recordings to below 50\% under all tested speech recognition systems, which is much better than existing solutions.

\end{abstract}


\section{Introduction}
\label{sec:intro}
\begin{figure}[t]
    \centering
    \includegraphics[width=\linewidth]{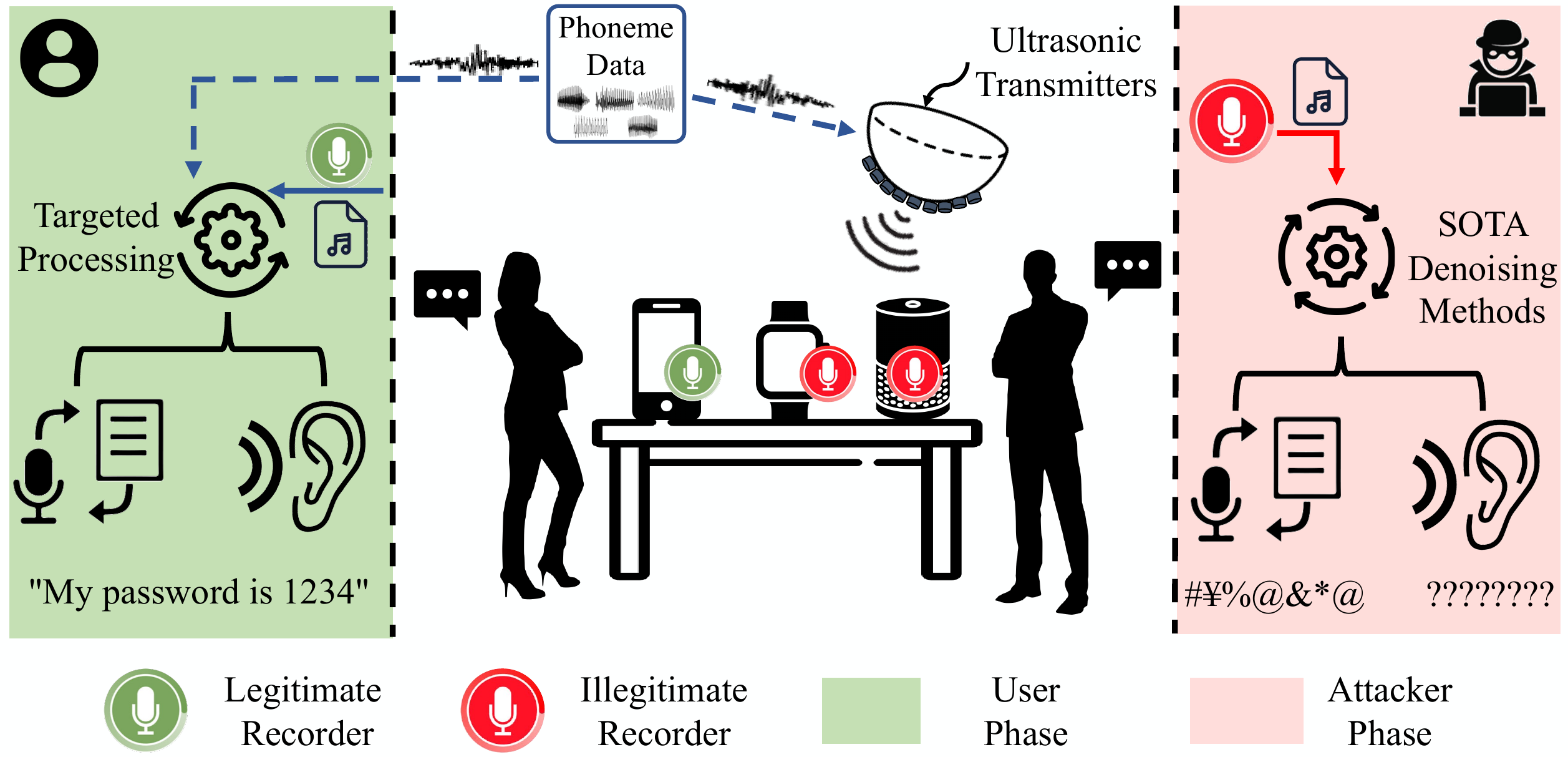}
    \caption{Anti-Eavesdropping with user-controlled recording.}
    \label{fig:introduction_system}
    \vspace{-7mm}
\end{figure}

\IEEEPARstart{M}{icrophones} are widely embedded in commercial electric device nowadays, which intensifies concerns over voice privacy. As shown in Figure~\ref{fig:introduction_system}, people are surrounded by various microphone-equipped devices in daily life, such as smartphones and smart speakers. Due to their blackbox nature, it is possible for adversaries to exploit, compromise, or even misconfigure these devices for eavesdropping. With the help of automatic speech recognition (ASR) systems, victims' personal data could be extracted thus violating their privacy. Many news reports the risk of being eavesdropped all the time by devices equipped with Siri, Alexa, or Google Assistant~\cite{android_police,alexa_google,siri}. Some reported cases also show severe consequences caused by eavesdropping:
In 2021, a leaked recording says that the Revolutionary Guards Corps overruling many government decisions, which puts the speaker in the recording, the Foreign Minister of Iranian, into a particular point of contention~\cite{iran_foreign_minister}. In 2020, the Ukraine prime minister submitted his resignation because a leaked recording suggesting he had criticized the president~\cite{GuardianUkraine}. In 2018, the defense secretary of the UK was interrupted by voice assistant during Commons statement because Siri listens constantly to seek the wake word~\cite{BBCSiri}, which can be treated as a new form of eavesdropping. 

The critical condition of voice privacy shows the necessity of anti-eavesdropping techniques. However, existing solutions fall far short of needs.
For example, Project Alias~\cite{alias} and Paranoid Home Wave~\cite{Paranoid} inject white/chatter noise into microphones to prevent potential sound monitoring. 
Nevertheless, to reduce the disturbance caused by the audible noise, users have to attach the noise transmitter to the microphone to lower the noise energy, which is not applicable in some scenarios.
GhostTalk~\cite{GhostTalk} applies electromagnetic interference (EMI) to induce noises to internal circuits of microphones, but the EMI could affect nearby devices like cardiac pacemakers, which could lead to unexpected results.

To overcome these difficulties, researchers attempt to jam recorders in other ways. 
In 2017, Roy and Zhang et al. reveal a new side channel in microphones that can inject audible signals into microphones with inaudible ultrasound~\cite{10.1145/3133956.3134052,backdoor}. The key point is to modulate the target audible signal on an ultrasonic carrier before sending it to the microphone. Once being received, the modulated signal will be demodulated automatically because of the non-linearity, as shown in Figure~\ref{fig:nonlinearity_in_microphone}. Its acceptable transmission distance ($>$2m) and neglectable influence on nearby users and devices make this method a promising solution for jamming voice recorders.

Based on this, Chen~et al. implement a bracelet-like wearable to emit ultrasound noises, achieving good jamming coverage by utilizing users' irregular arm movements~\cite{CHI}. Li~et al. propose Patronus~\cite{patronus} that can disturb recorders with carefully designed noises and adopts an iron reflection layer to extend coverage. Gao~et al. design MicFrozen~\cite{gao_cancelling_2023}, which reduces signal-to-noise ratio (SNR) of recordings by cancelling voice signals and adding coherent noises. Sun~et al. design MicShield~\cite{sun_alexa_2020} which provides selective jamming for smart speakers based on wake word detection.

We systematically analyze existing studies and reveal their limitations, as shown in Table~\ref{tab:overview}. First, most of them adopt simple-form noises such as White Gaussian Noise (WGN) and random frequency noise, which achieve audio jamming solely based on high power. On the one hand, the high power ultrasound could harm the human ear~\cite{human_ultrasound_tolerance}. On the other hand, the non-linearity also exists in transmitters~\cite{211283}, so the energy of the transmitted noise should be controlled within a certain range to avoid emitting audible noises. These two issues create a central dilemma between jamming coverage and usability. Besides, these simple-form noises can be easily removed by noise reduction techniques, thus cannot prevent the leakage of privacy thoroughly~\cite{Walker2021Evaluating}. For the works utilizing speech-based noise~\cite{gao_cancelling_2023}, they employ continuous speech signals, which gives a chance to remove them from recordings with speech separation techniques. This fact is devastating for an anti-eavesdropping application as an adversary is very likely to apply state-of-the-art (SOTA) denoising techniques or even targeted noise removal methods before privacy extraction. In addition, most of existing works do not support authorized recording, which limits the usage scenario.

\begin{table}[]
    \caption{An Overview of Anti-Eavesdropping Systems}
    \resizebox{\columnwidth}{!}{
        \begin{threeparttable}
        \begin{tabular}{cccccc}
            \toprule
            System     & Noise Form    &\begin{tabular}[c]{@{}c@{}}Commercial\\ ASRs~$\dagger$\end{tabular}  & \begin{tabular}[c]{@{}c@{}}SOTA \\ Enh.$~\dagger$\end{tabular}& \begin{tabular}[c]{@{}c@{}}Specialized\\ Enh.$~\dagger$\end{tabular} & \begin{tabular}[c]{@{}c@{}}Recording\\Support\end{tabular}\\ \midrule

            MicFrozen~\cite{gao_cancelling_2023}   & Speech~*  & \ding{51}             & \ding{51}       & \ding{55} & \ding{55}               \\
            MicShield~\cite{sun_alexa_2020}     & WGN           & \ding{51}             & \ding{55}         &      \ding{55}& \ding{55}        \\
            Roy et al.~\cite{backdoor} & WGN~*     & \ding{55}              & \ding{55}         & \ding{55}    &  \ding{55}   \\
            Patronus~\cite{patronus}   & Random Freq.~*  & \ding{51}             & \ding{55}         & \ding{51}    & \ding{51}          \\
            Chen et al.~\cite{CHI}      & WGN           & \ding{51}             & \ding{51}       & \ding{55}         &  \ding{55}    \\
            This work     & Phoneme~* & \ding{51}             & \ding{51}       & \ding{51}     & \ding{51}         \\ \bottomrule
        
        \end{tabular}

        \begin{tablenotes}
            \small 
            \item \textit{Note: * in Noise Form means generated based-on, not exactly the same. \ding{55} in $\dagger$ columns means that the authors have not conducted relevant experiments.}
            
        \end{tablenotes}
    \end{threeparttable}
    }
    \label{tab:overview}
    \vspace{-6mm}

\end{table}

In this paper, we propose an anti-eavesdropping system that achieves effective and reliable jamming with support for authorized recording in real-world privacy-preserving scenarios. Instead of solely relying on high power, we explore the idea of informational masking, which achieves jamming by disturbing the signal structure in recordings. Specifically, we design a new type of jamming noise that contains multiple phoneme sequences. When jammed by our noise, the phoneme structure of the speech signal would be greatly obscured. Since phonemes are the basic units of sound for distinguishing words, the chaotic phoneme pattern makes the speech signals hard to understand. Moreover, our noise shows inherent robustness against both existing SOTA and targeted-trained denoising methods. The discrete phonemes in our noise make denoising methods fail to disentangle the genuine speech elements from distracting ones due to their high similarity. We encourage readers to listen to audio examples at the demo website~\footnote{\url{https://github.com/desperado1999/InfoMasker}. Relevant codes will also be released here.}. The utilization of information masking also relaxed the requirement on high energy, and we further address the challenge on increasing the jamming range without losing inaudibility. At last, we propose a transformer-based content recovery method to provide support for authorized recordings.

In general, we summarize our contributions as follows:
\begin{itemize}
    \item We propose a new type of jamming noise, named Phoneme-Based Audio Jamming Noise, based on the idea of informational masking.
    \item We design and implement a system based on our noise, which can prevent eavesdropping while preserving recording permission for authorized users.
    \item We conduct comprehensive experiments to evaluate our system from different aspects. The results prove the high effectiveness and robustness of our system and show its usability in real-world scenarios.
\end{itemize}
\section{Preliminary}
\label{sec:preliminary}


In this section, we introduce the nonlinearity effect in the microphone, then we describe the linguistic structure of speeches and how humans and machines understand them.

\subsection{Nonlinearity in Microphone}
\begin{figure}[t]
    \centering
    \begin{subfigure}[b]{0.49\columnwidth}
        \centering
        \includegraphics[width=\linewidth]{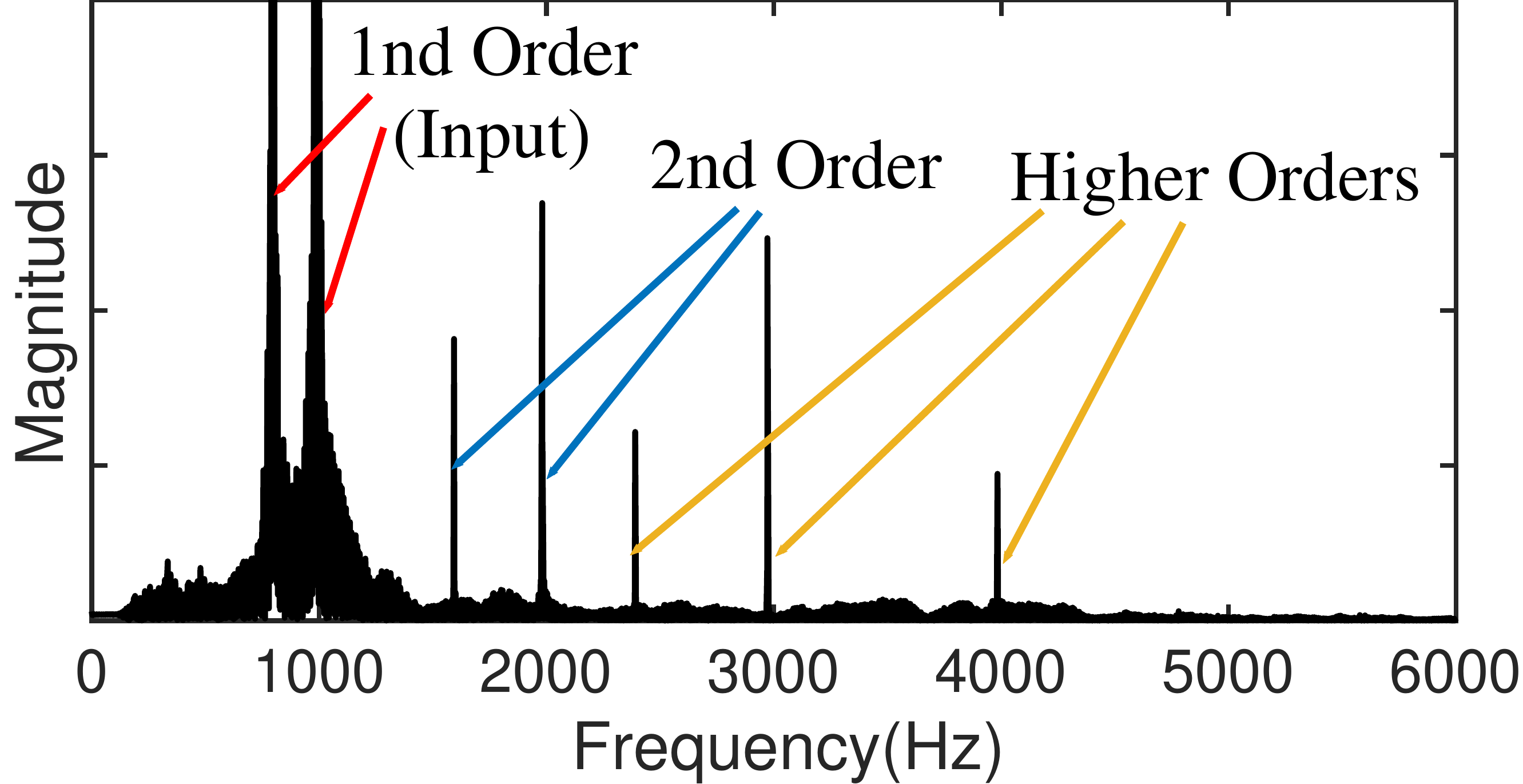}
    \end{subfigure}
    \begin{subfigure}[b]{0.49\columnwidth}
        \centering
        \includegraphics[width=\linewidth]{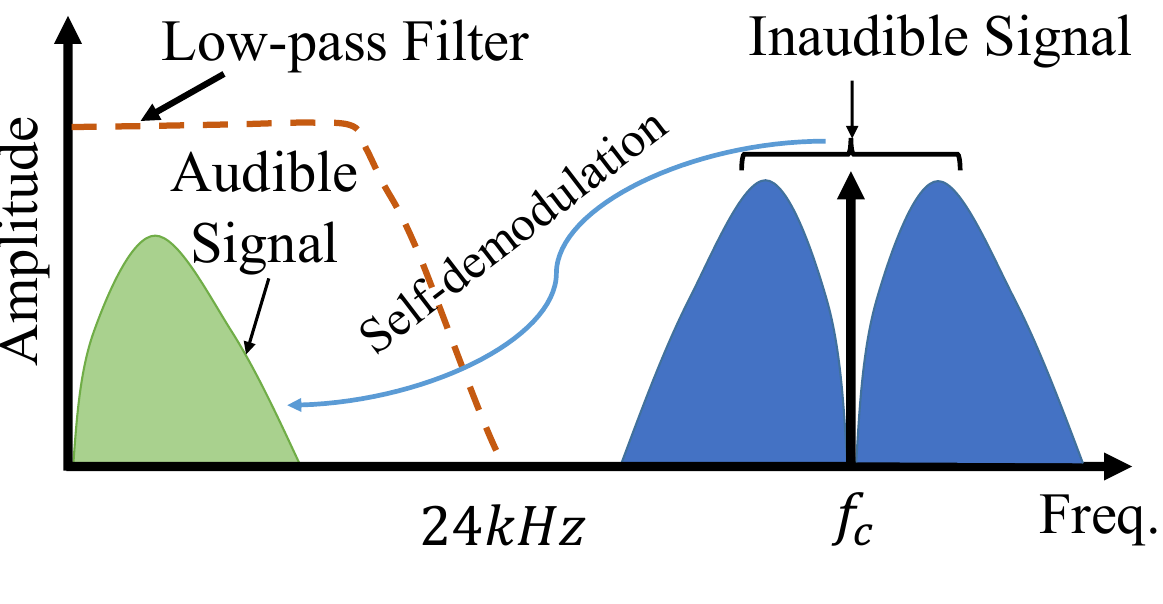}
    \end{subfigure}
    \caption{Nonlinearity in microphone. The left figure shows the audio spectrum recorded by a Huawei P10 smartphone with an input of two single tones (800Hz and 1000Hz).}

    \label{fig:nonlinearity_in_microphone}
    \vspace{-5mm}
\end{figure}

A microphone is a type of transducer that converts acoustic signals into electrical signals. Previous studies have shown that the preamplifier in most types of microphones, including Electret Condenser Microphones (ECMs) and Micro Electro Mechanical Systems (MEMS) Microphones, involves nonlinear operations that cause inter-modulation distortion in its output \cite{7569908}. As a result, the microphone's output contains both frequency components of the input and all possible linear combinations of them \cite{zi-qiang_lang_evaluation_2000}. To illustrate, suppose the inputs are two single tones with frequencies of $f_1$ and $f_2$, the nonlinearity makes the output contain not only components with frequencies of $f_1, f_2$, but also $f_1+f_2$, $f_1-f_2$, $2f_1$, $2f_2$, ... etc., as shown in the left of Figure~\ref{fig:nonlinearity_in_microphone}.

To inject an audible noise signal $n(t)$ into a microphone stealthily, we first modulate $n(t)$ on a high-frequency carrier $c(t)$ = $cos(2\pi f_c t)$ via amplitude modulation and then transmit the modulated signal along with the carrier simultaneously. When these signals arrive at a microphone, the non-linearity produces distortion that are harmonics and cross-products of the carrier and the modulated noise~\cite{10.1145/3133956.3134052}, generating a low-frequency shadow signal and other high frequency components. The shadow signal is the same as $n(t)$ and other components will be filtered out by the low-pass filter in the microphone, as shown Figure~\ref{fig:nonlinearity_in_microphone} (right part).

\subsection{Informational Masking}
\label{subsec:informational_masking}
Informational masking, which is first defined in~\cite{Pollack1975-qm}, describes the degradation of the auditory detection threshold in the human brain when the target sound is embedded in other interferers with similar characteristics. Informational masking is usually associated with its complementary term: energetic masking, which occurs when interferers are present at the same time and the same frequency bands~\cite{leek_informational_1991}. Unlike energetic masking which mainly depends on the relative energy between the target and the interferer in each frequency band, the degree of information masking mainly depends on the similarity between the target and the interferer. Generally speaking, these two types of masking are not independent and they always affect the auditory detection threshold simultaneously.

\subsection{Human Auditory System and ASR}
\label{subsec:human_auditory_system_and_asr}
One of the main tasks of the human auditory and ASR systems is extracting semantic information from speech signals. In order to improve speech intelligibility, both systems need to first eliminate the noise in the signals, then extract phoneme series, the primary component of a speech signal, and decode the phoneme sequence into meaningful content. 

The attention mechanism in the human auditory system, as known as "Cocktail Party Effect", has a strong noise reduction ability~\cite{festen_effects_1990, leek_informational_1991,brungart_informational_2001, mattys_effects_2010}. It allows a listener to distinguish signals from other sources and then eliminates the influence of uninterested noises, thus achieving denoising. To be more vivid, imagine you are whispering to a friend in a crowded cocktail bar where many people are talking loudly, although other people's voices may be louder than your friend's, you can still focus on his/her voice and understand what he/she is talking about. This mechanism also helps human beings to reduce the impact of masking effect caused by interferences. The effectiveness of the attention mechanism mainly depends on the degree of difference between the target signal and the noise in three aspects: fundamental frequency, temporal properties, and spatial distribution. A larger difference means better discrimination with the help of the attention mechanism.

A typical ASR system works similarly to the human auditory system to "comprehend" speech signals. Most of these ASR systems will first extract voice features, such as mel-frequency cepstral coefficients (MFCC), from the audio segments and then recognize the phoneme series from these features using an acoustic model. Then with the help of the pronunciation model and language model, the phoneme series will be decoded into normal text. To improve recognition accuracy, speech enhancement methods are always applied before recognition. 
The widely used speech enhancement methods can be roughly divided into two types: noise reduction and noise separation. The former, such as spectral subtraction, targets on reducing the noise from the signals. This type of method relies on the invariance of statistical characteristics of noise and the differences in temporal properties between the noise and the target speech signal. The noise separation, such as blind signal separation, targets on separating source signals from mixed signals. Such methods rely on the assumption that each source signal is independent. 

In this work, we aim to prevent both humans and machines from extracting semantic information embedded in speech signals by injecting noise signals into the recordings. To achieve this target, the injected noise should interfere with human and ASRs' understanding of semantics. Equally critical, the injected noise should be difficult to remove by both auditory attention mechanism and speech enhancement algorithms. Therefore, we design our noise that is highly correlated with speech signals to affect the phoneme structure of the target speech signal, which not only disturbs human's and ASR's understanding of the semantics, but also improves the robustness of noise against speech enhancement methods.

\section{Problem Formulation}
\label{sec:problem_formulation}
In this section, we first introduce the system and threat model, and then discuss the design goals of our system.

\subsection{System Model}
\label{subsec:system_model}

We consider the scenario where the users are having a conversation in an indoor environment such as an office or a meeting room. The users may need to record the conversation while they also want to prevent the conversation from being eavesdropped by possible unknown adversaries. As shown in Figure \ref{fig:system_model}, the system involves four entities:

\begin{figure}[t]
    \centering
    \includegraphics[width=\linewidth]{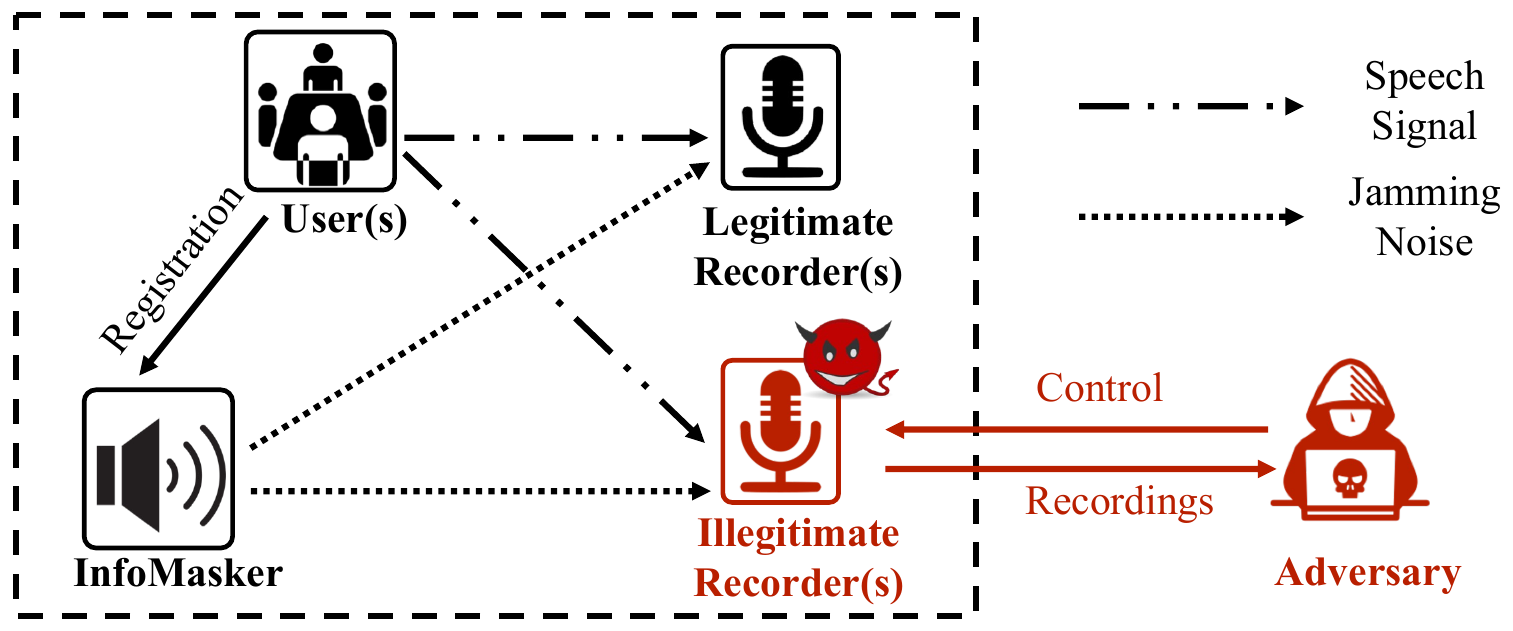}
    \caption{System Model.}
    \label{fig:system_model}
    \vspace{-7mm}
\end{figure}

\textbf{User(s) }are the people who are having a conversation. To safeguard their privacy, they deploy \textit{InfoMasker} to prevent their voice from being eavesdropped.

\textbf{InfoMasker }is a user-controlled jamming device which can generate and emit jamming noise according to the users' registration information to prevent possible eavesdropping. Please note that we do not cover the voice call scenario since all microphones in the environment will be jammed.


\textbf{Legitimate Recorder(s) }are placed by the users to record the ongoing conversation. Because of the presence of InfoMasker, the recorded audios will also be affected and will contain both the conversation and the jamming noise. After the conversation, the users will try to recover the content from the noisy recording. Comparing to the adversary, the users know the detail of the injected noise.

\textbf{Illegitimate Recorder(s) }refer to all other devices equipped with microphones in the environment, such as smartphones and smart watches. Due to the black box nature of most electronic products, they are not fully controlled by users and may eavesdrop on users secretly. Considering the ease of use, users are unlikely to put these devices in hidden areas which may cause non-line-of-sight (NLOS), so here we do not consider the NLOS scenario.

\subsection{Threat Model}
We consider an adversary who can control recording devices in the environment (e.g., the manufacturer of the smart home devices). To eavesdrop on the content of the conversation, the adversary would record the speech signals, improve the recordings' intelligibility using various methods, and then extract semantic information. In this work, we assume the adversary has the following capabilities in each step of eavesdropping:

\textbf{Audio Recording. }We assume the adversary can control one or more recording devices in the environment to record single-channel or multi-channel audio signals.

\textbf{Speech Enhancement.}
We assume that the adversary can improve intelligibility of the recorded speech signals with different speech enhancement methods, including noise reduction, speech separation, and Blind Signal Separation (BSS) if multi-channel recordings are acquired.

\textbf{Semantic Information Extraction. }We consider that the adversary could jointly use different ASRs together with human ear to extract semantic information from the enhanced recordings, where human can recognize recordings accurately and ASRs can interpret speech contents efficiently. 

We also consider a powerful adversary who knows details of our noise generation method and could train a customized ASR system targeting on our noise.

\subsection{Design Goals}
We envision the following design goals of InfoMasker:

\textbf{Effectiveness}: 
Noise transmitted by InfoMasker should be able to prevent both ASR systems and the human auditory system from recognizing the noisy recording, which contains voices from single or multiple users.

\textbf{Robustness}: Noise injected by InfoMasker should be robust against various speech enhancement methods.

\textbf{Low-interference}: Noise signal transmitted by InfoMasker should be inaudible and harmless to human beings.

\textbf{Controlled Recording Privilege}: Users with the recording privilege can recover the content from noisy recordings.

\section{Phoneme-Based Informational Masking}

\begin{figure}[t]
    \centering
    \includegraphics[width=\linewidth]{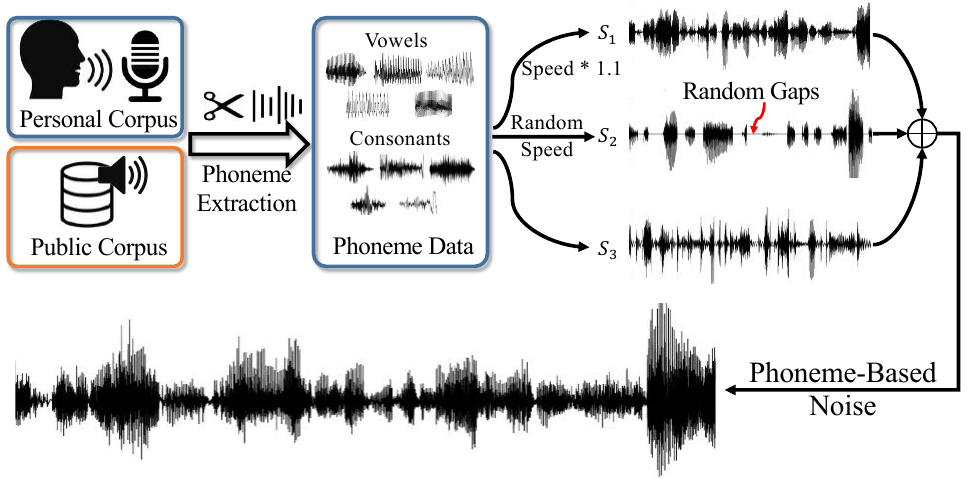}
    \caption{Generation of phoneme-based noise. }
    \label{fig:full_noise_design}
    \vspace{-6mm}
\end{figure}

\label{sec:noise_design}
\subsection{Key Insight}
The main goal of this paper is jamming microphones in the environment by transmitting noise to make the recordings unrecognizable to both human and ASR systems. The success of jamming ASR systems mainly relies on the noise's robustness against speech enhancement methods, while the success of jamming the human auditory system mainly relies on the two masking effects: informational masking and energetic masking. The effectiveness of energetic masking mainly depends on the relative energy of the noise to the target signal, while for the latter, it mainly depends on the content of the noise~\cite{brungart_informational_2001,leek_informational_1991, festen_effects_1990}. Existing works \cite{backdoor, patronus, CHI} focus on utilizing energetic masking, which leads to a high demand for energy and is not suitable in the ultrasonic transmission scenario.

In this paper, our key insight is exploiting the informational masking effect to form noise. Previous studies have shown that when presented with noises, the differences in phonetical properties such as fundamental frequency (F0) between the noise and the target speech will assist listeners to filter out the noise and understand the target speech signal\cite{brungart_informational_2001}. Therefore, we first construct noise that has similar F0 to the target speech signal. Additionally, through experiments we find that the difference in speech rate between the noise and the speech also affects the effectiveness of informational masking, which inspires us to treat speech rate as another factor in our phoneme-based noise design.

\subsection{Noise Design}
\label{subsec:noise_design}

As shown in Figure \ref{fig:full_noise_design}, our phoneme-based noise consists of three phoneme sequences: $S_1$, $S_2$, and $S_3$. $S_1$ is a sequence of random vowels without gaps in between, and it is accelerated by a preset parameter to include more vowels per unit time for better jamming effectiveness, and the parameter is set to 1.1 according to our experiments. The reason we choose vowels is that compared to consonants, vowels make up most of the energy in speech signals. Because the difference in phonetical properties such as fundamental frequency (F0) between the speech signal and the noise will assist listeners in separating the noise, we extract the phoneme data from the target people's speech materials to minimize such difference. Simple concatenation of the phoneme data will cause discontinuity at phoneme boundaries. To smooth the connected phoneme data, we apply a hamming window with a length of 25ms at the juncture between phonemes. Without special mention, this smoothing method is also applied to the following sequences that make up our noises.

Then we consider narrowing down the speech rate gap between the noise and the target signal. Since the target signal's speech rate is unknown, we add another vowel sequence $S_2$ with random speech rate. Different from $S_1$, there are gaps of random length between the vowels in $S_2$. Each vowel is accelerated or decelerated with a randomly chosen speed factor that follows a uniform distribution $U(0.3, 1.8)$ which is chosen according to our experimental results. In addition to making the rate more similar to the target signal, $S_2$ also makes the noise more varied and shows less constant patterns, which helps with the robustness against noise reduction methods.

Apart from vowels, we add a consonant sequence $S_3$ to increase the diversity of noise. As the difference in consonants between different people is much smaller than that in vowels, we utilize the consonant data from a public speech corpus (LibriSpeech \cite{LibriSpeech}). Then we connect these consonants without gaps to form $S_3$.

\begin{figure}[!t]
    \centering
    \begin{subfigure}[b]{0.49\columnwidth}
        \centering
        \includegraphics[width=\linewidth]{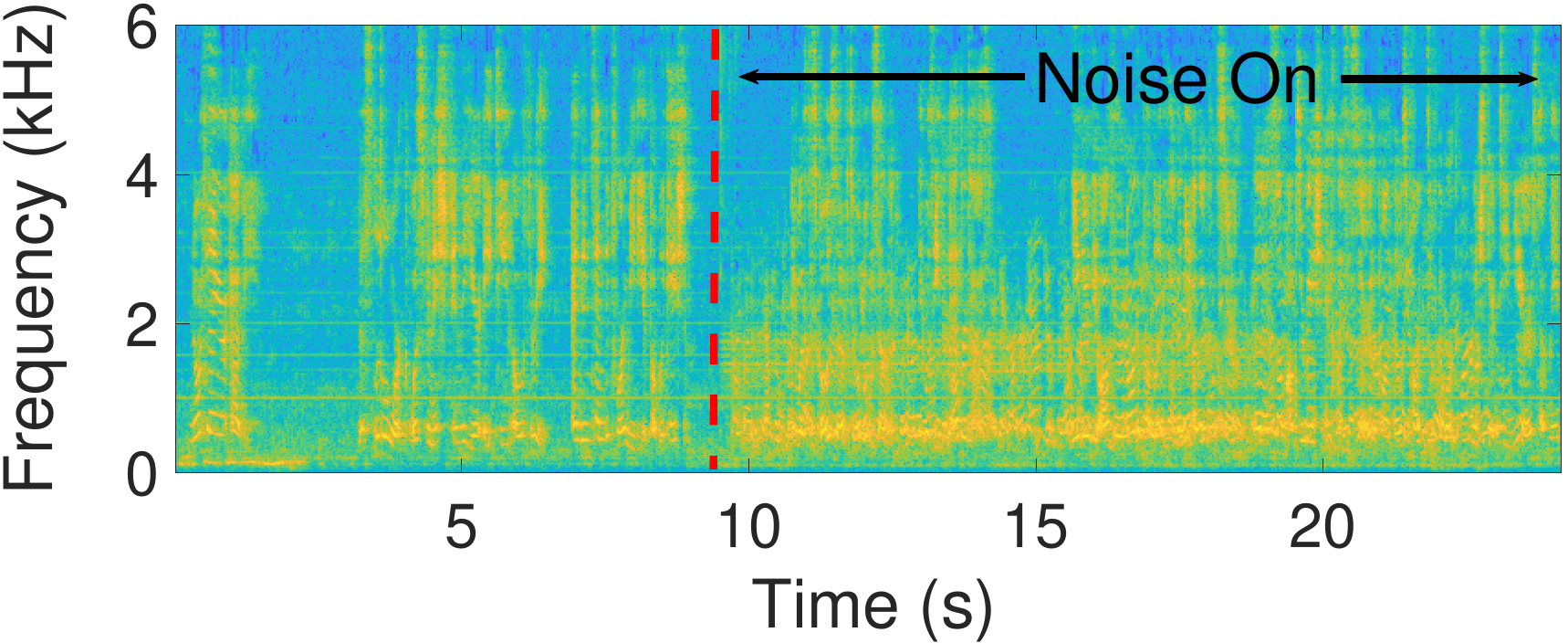}
        \caption{Phoneme-Based Noise}    
        \label{fig:designed_denoise}
    \end{subfigure}
    \begin{subfigure}[b]{0.49\columnwidth}  
        \centering 
        \includegraphics[width=\linewidth]{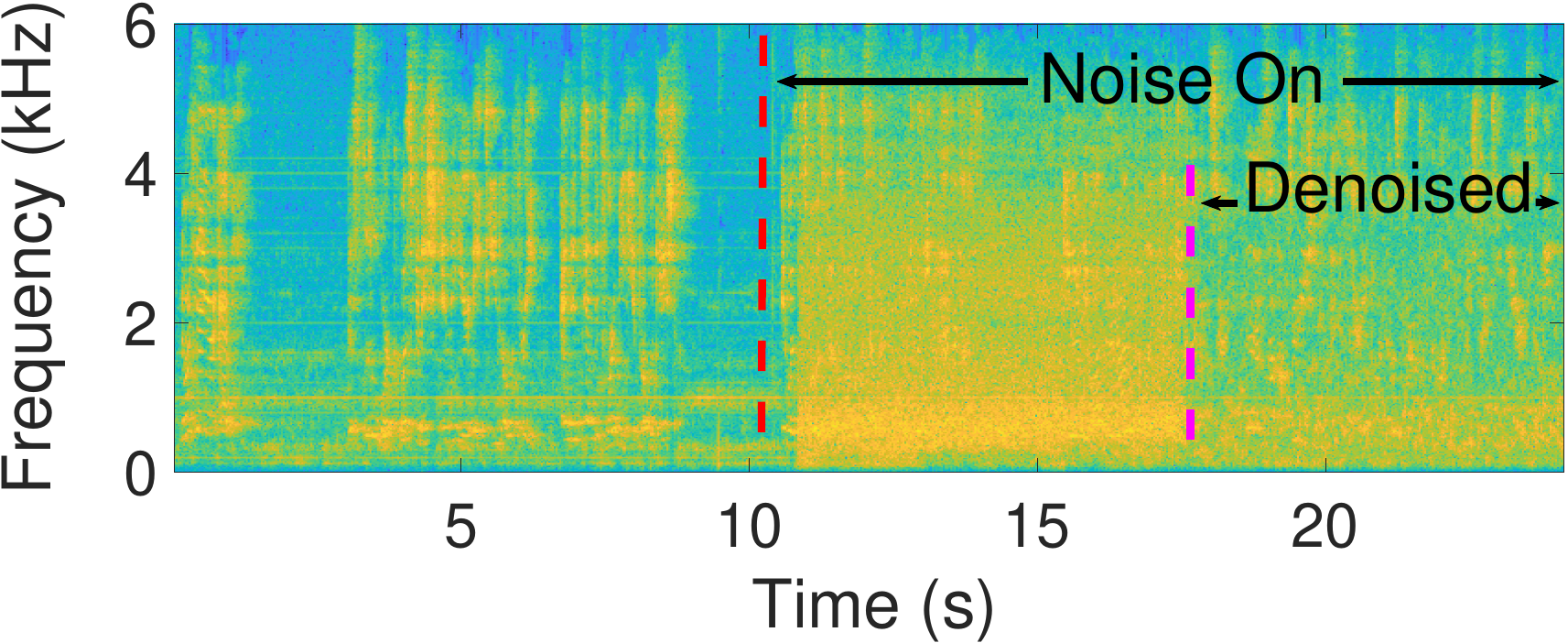}
        \caption{White Noise}    
        \label{fig:white_denoise}
    \end{subfigure}

    \begin{subfigure}[b]{0.49\columnwidth}   
        \centering 
        \includegraphics[width=\linewidth]{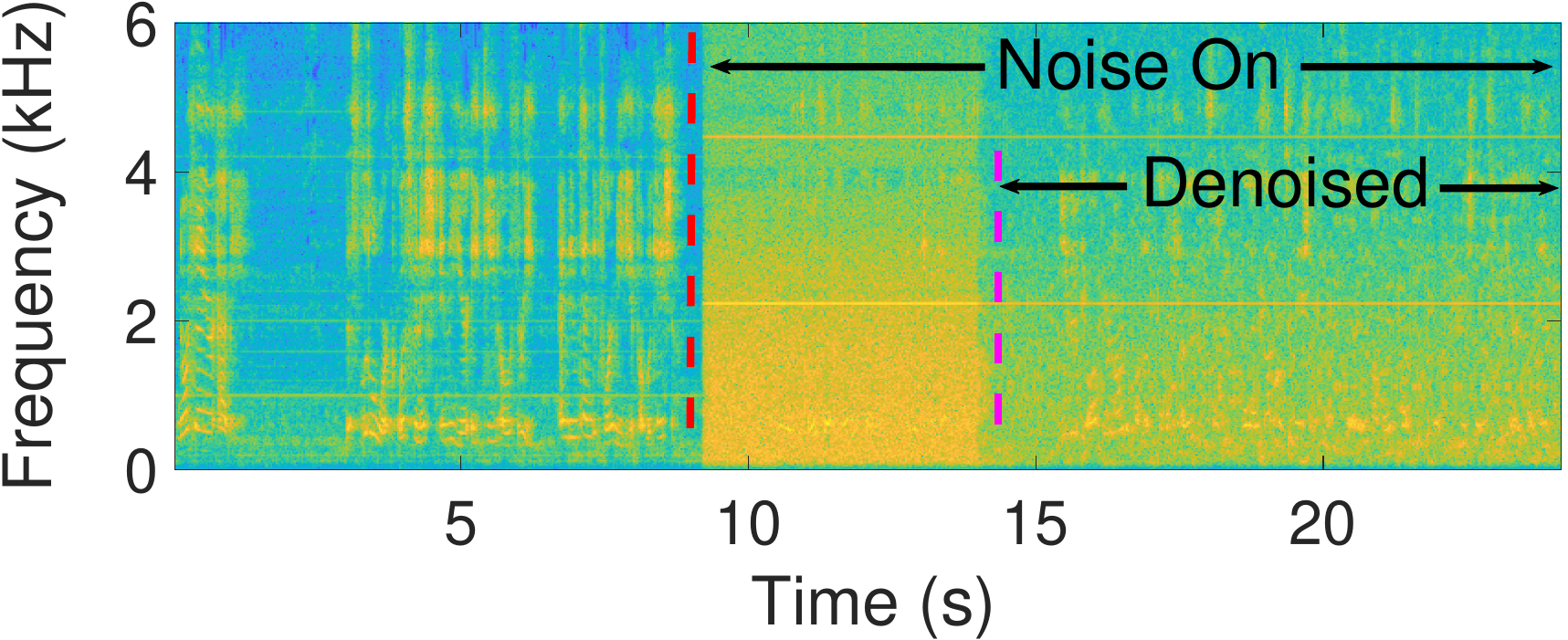}
        \caption{Noise in \cite{CHI}}
        \label{fig:chi_denoise}
    \end{subfigure}
    \begin{subfigure}[b]{0.49\columnwidth}   
        \centering 
        \includegraphics[width=\linewidth]{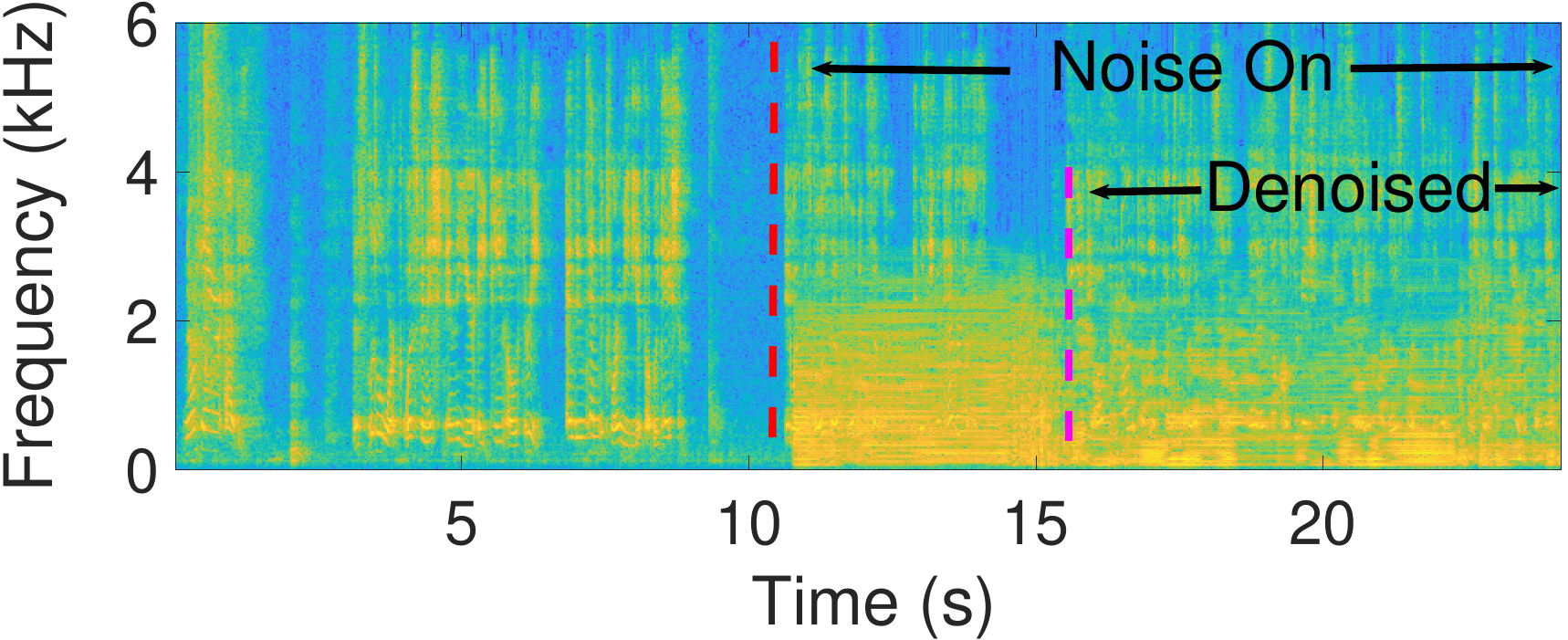}
        \caption{Noise Generated by \cite{commercial_device}}
        \label{fig:commercial_denoise}
    \end{subfigure}

    \caption{Comparison of robustness of different noises against the built-in noise reduction in a Vivo Nex smartphone.}
    \label{fig:test_denoise_in_noise_design}
    \vspace{-6mm}
\end{figure}

Our noise is the superposition of the above three phoneme sequences. To demonstrate the feasibility, we test its robustness against the built-in noise reduction in a smartphone. We play voices along with various types of noise and record the signal with the phone. The results in Figure~\ref{fig:test_denoise_in_noise_design} show that our noise is robust against the built-in noise reduction while the others are suppressed significantly.

\begin{figure*}[t]
    \centering
    \includegraphics[width=\linewidth]{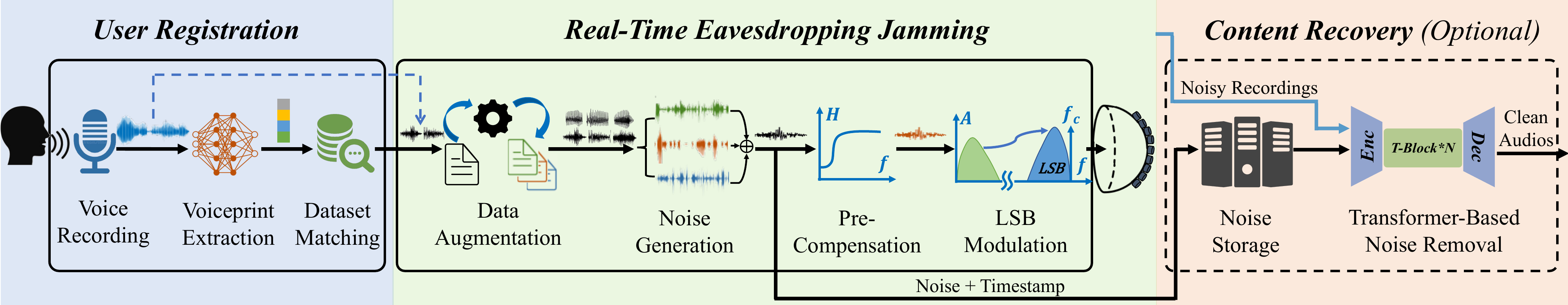}
    \caption{Workflow of InfoMasker.}
    \label{fig:system_design}
    \vspace{-7mm}
\end{figure*}

\section{System Design}
\label{sec:system_design}

\subsection{System Workflow}

The workflow of InfoMasker is shown in Figure~\ref{fig:system_design}, involving three phases: \textit{User Registration}, \textit{Real-Time Eavesdropping Jamming}, and \textit{Content Recovery}. In the first, InfoMasker collects speech materials from user(s) and extracts their voiceprints. Depending on the amount of materials, it adopts different methods to generate adequate phoneme data.

The phoneme data is then used for noise generation in the second phase, in which InfoMasker continuously generates, compensates, modulates, and transmits the jamming noise to jam microphones in the environment.

If recording privilege is required, the noise will be stored with timestamps locally and fed into a denosing network together with noisy recordings to get clean audios.

\subsection{User Registration}
\label{subsec:user_registration}

As stated in Section \ref{sec:noise_design}, the phoneme-based noise is generated according to the user's speech data. However, collecting a large number of speech samples of the user is time-consuming thus sometimes not practical. Our solution is to pick speech data similar to the user's voices from a dataset. The similarity between the voices and the dataset audios is represented by the cosine similarity of their voiceprints, which are extracted with the method proposed in~\cite{GE2E}. Denote two voiceprints as $\bm{e_1}$ and $\bm{e_2}$, the cosine similarity of them can be represented as $1-d(\bm{e_1}, \bm{e_2})$, where $d(\bm{e_1}, \bm{e_2}) = \frac{\bm{e_1}\cdot \bm{e_2}}{||\bm{e_1}||||\bm{e_2}||}$.

\begin{figure}[t]
    \centering
    \begin{minipage}[t]{0.49\linewidth}
        \centering
        \includegraphics[width=\linewidth]{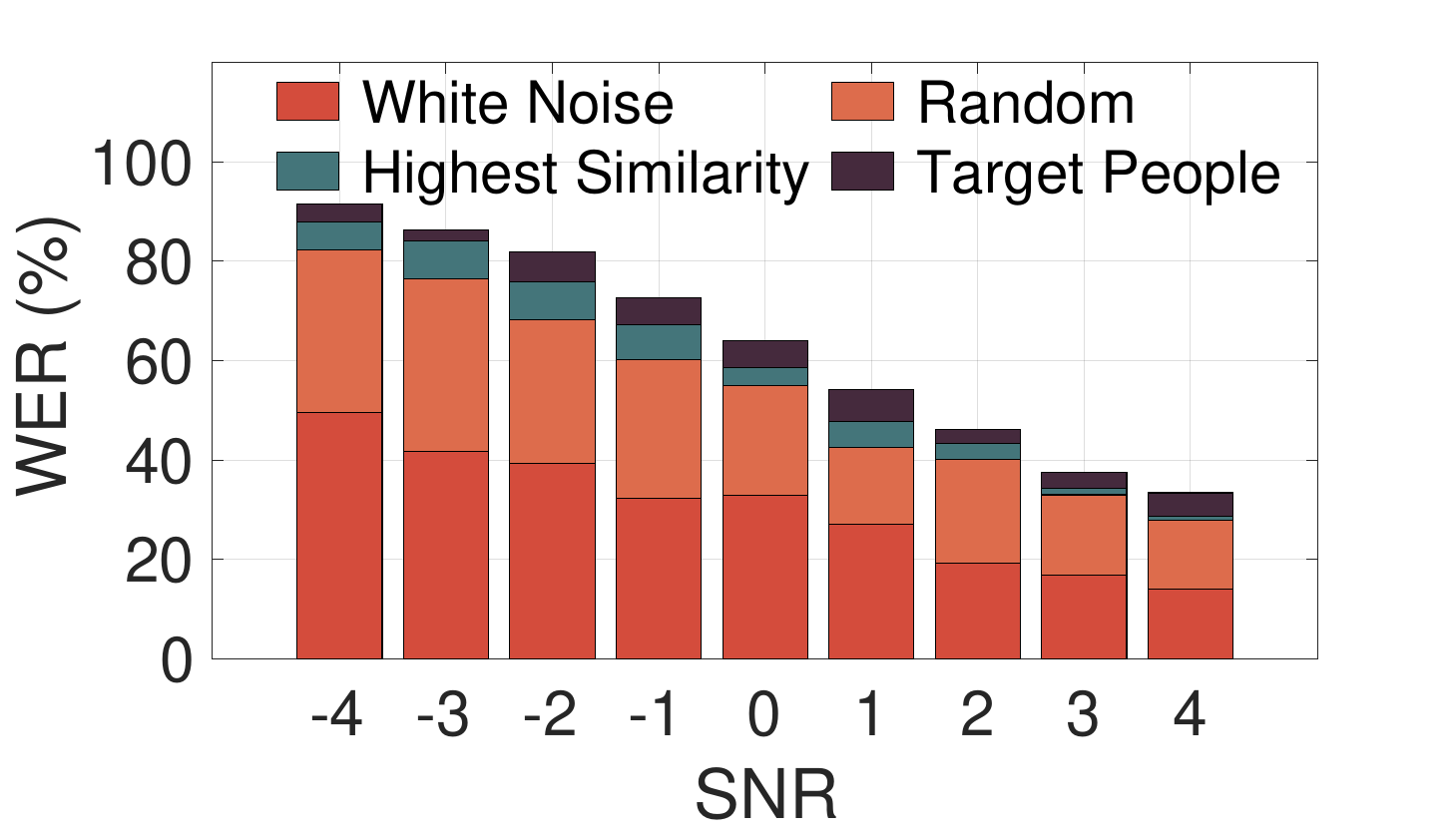}
        \caption{Effectiveness of different types of noise.}
        \label{fig:jamming_with_other_data}
    \end{minipage}
    \begin{minipage}[t]{0.49\linewidth}
        \centering
        \includegraphics[width=\linewidth]{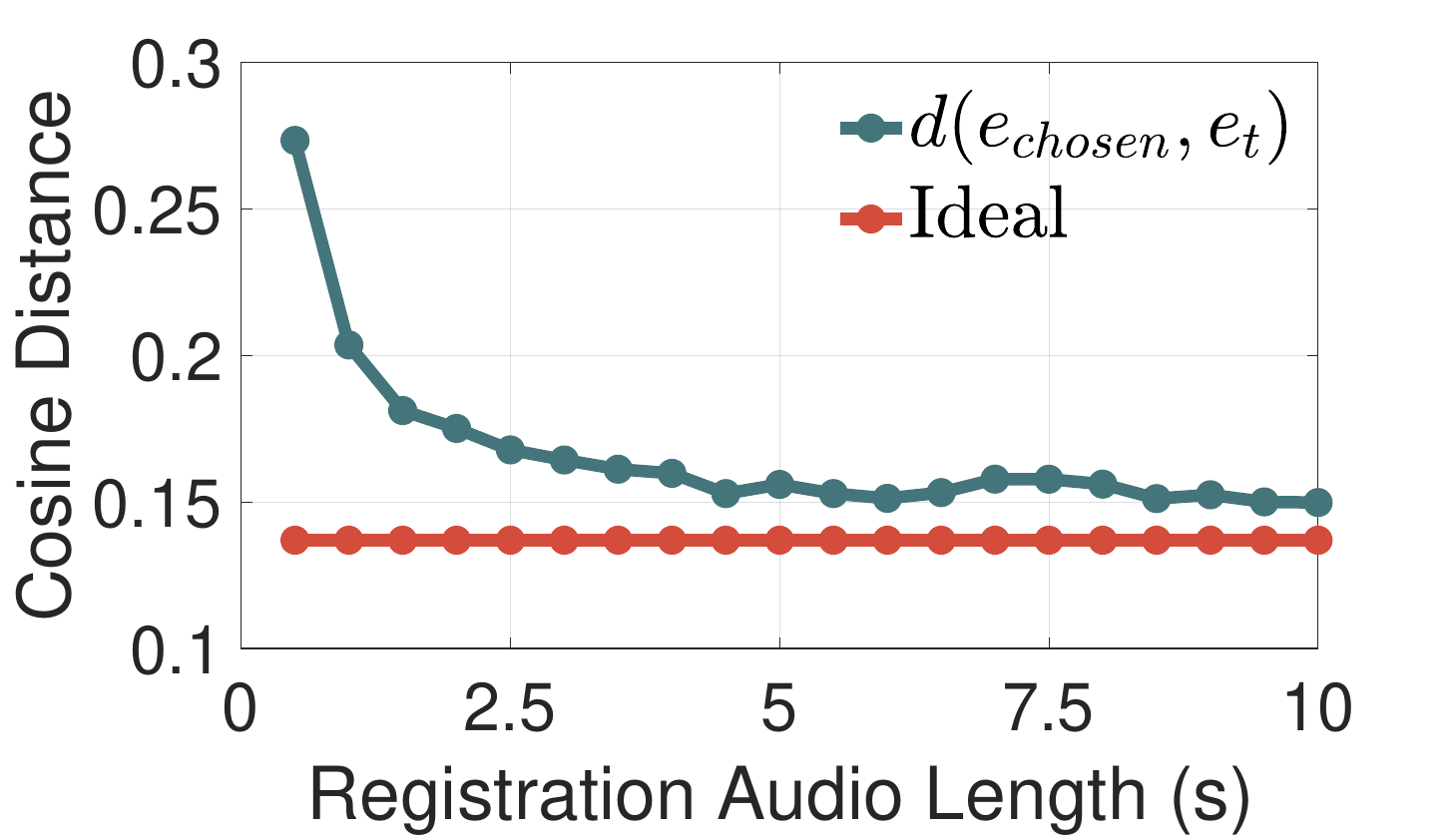}
        \caption{Comparison of different registration speech length}
        \label{fig:test_registration_length}
    \end{minipage}
    \vspace{-5mm}
\end{figure}

To evaluate the effectiveness of this solution, we test three types of noise generated from audios showing different similarities to the user's voice: the user's voice, the audio with the highest similarity in a dataset, and the randomly chosen audios from a dataset. We also use WGN as a baseline. The dataset we used here is the train-clean-360 subset from LibriSpeech~\cite{LibriSpeech}. As shown in Figure~\ref{fig:jamming_with_other_data}, the jamming performance of the noises decreases as the voice similarity drops, and there is an about 10\% gap in Word Error Rate (WER) between the best and the worst case~\footnote[1]{In this paper we use Tencent ASR \cite{tencent_asr} for speech recognition if there is no special description}. Based on these results, we consider two types of registration scenarios according to the number of users.

\textbf{Single-User Registration. }
In this scenario, our priority is protecting a specific user's voice privacy. Meanwhile, the others' voices are also protected, merely with a 10\% drop in effectiveness as shown in Figure~\ref{fig:jamming_with_other_data}

There are two registration ways depending on the user's time availability. Given ample time, the user can record adequate speech data using Harvard Sentences dataset~\cite{harvard_sentences}, which contains phonetically balanced sentences. These recordings can provide sufficient and balanced phoneme data for noise generation. In contrast, when time is limited, the user can choose to record a few sentences for voiceprint extraction. Then the data with the most similar voiceprint in the dataset will be chosen for noise generation. 

To get the appropriate voice length for this time-restricted scenario, we conduct experiments to assess the impact of the voice length on the voiceprint similarity between the chosen data and the target people's voice (i.e., $d(\bm{e_{chosen}}, \bm{e_t})$). The results are shown in Figure~\ref{fig:test_registration_length} and the Ideal represents unrestricted voice length (utilizing all available data). We find that when voice length $<$ 5 seconds, the distance first drops rapidly as the length increases, and then drops much slower. The change in distance is neglectable when the length $>$ 5 seconds, indicating a 5-second speech is suitable for registration.

\textbf{Multi-User Registration. }
In this scenario, we protect the voice privacy of all the people present in the environment. Similar to the time-restricted scenario for single user, the users need to record speech data with Harvard dataset for voiceprint extraction. As each user's voiceprint is different, here we use the average of each user's voiceprint as the representative feature for this group of people, then finds the best-matching data in the dataset for noise generation.

Please note that InfoMasker is totally offline during the usage, so the registration will not cause privacy leakage.

\subsection{Real-Time Eavesdropping Jamming}
After the registration, when InfoMasker is turned on to protect users' voice privacy, it will continuously augment the phoneme data, generate jamming noise based on these augmented data, and then emit the noise to the air to jam possible eavesdropping devices.

\textbf{Data Augmentation. }
With the data generated from the registration phase, we are now able to generate phoneme-based noise. However, since the amount of phoneme data is limited by the dataset, the probability of occurring recurring fragments in the noise will increase as more noise is generated, especially when the dataset is small. When an attacker locates these fragments, it is very possible for him to recover part of the semantic information from the recordings with the help of language models, similar to the key reused scenario in running key cipher. To prevent this, a data augmentation process that fine-tunes the phoneme data is applied to increase the total amount of data. To retain a high similarity between the augmented data and the original voices, the fine-tune should be restricted to a people's inner differences. Here we adopt the following properties used in emotion recognition~\cite{speech_emotion_detection}: speech rate, F0 mean, F0 contour, and energy, which have inherent inner differences because of people's different emotions, as shown in Table~\ref{tab:phonetical_properties}. In addition, we randomly reverse each single phoneme data, which has little effect on the data's phonetical properties but disrupts its semantic information.

\begin{figure}[t]
    \centering
    \includegraphics*[width=\linewidth]{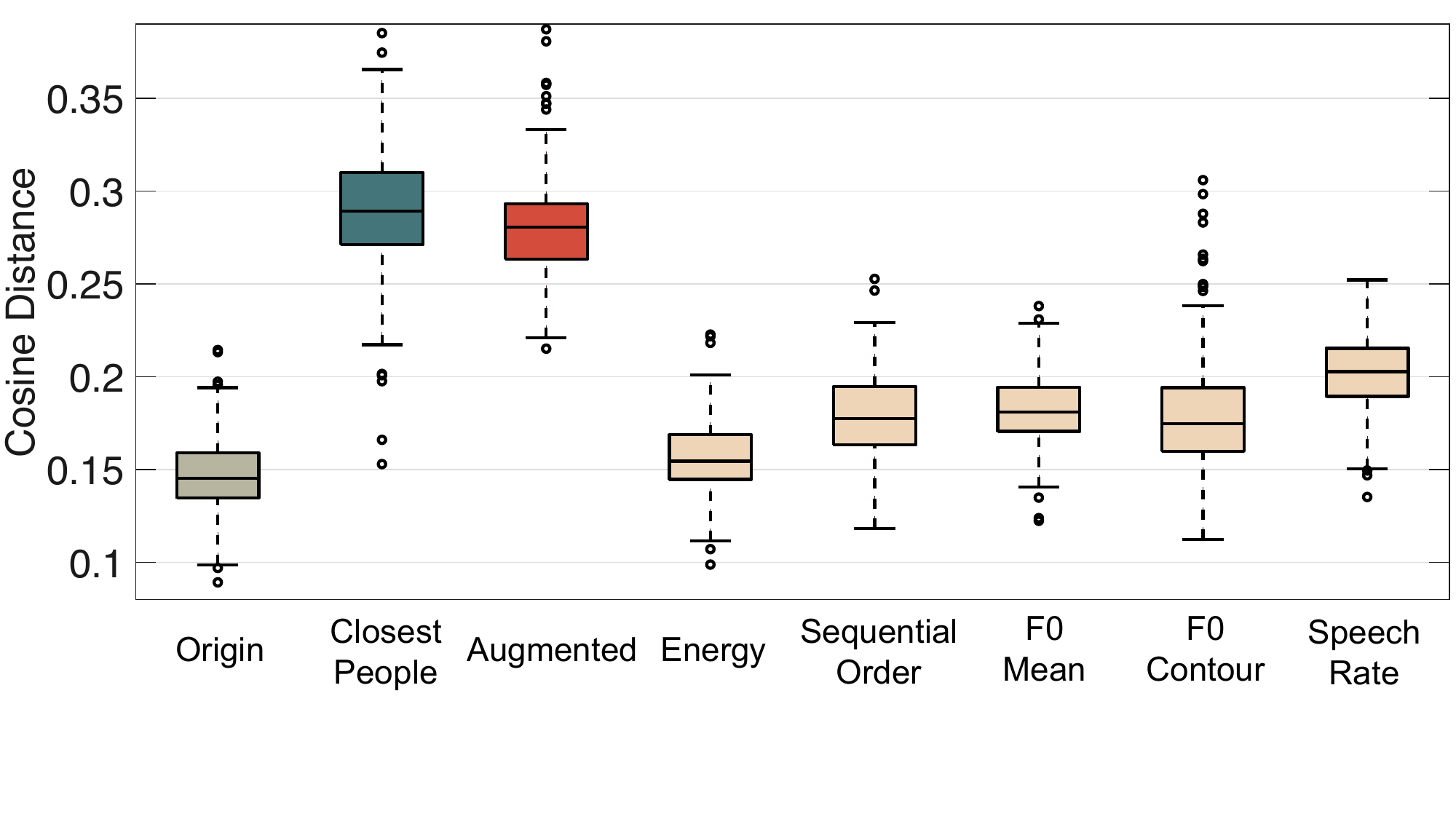}    
    \caption{Impact of data augmentation on the audio's phonetical features. The boxes illustrate the distribution of $d(\bm{\hat{e}}, \bm{e_t})$ where $\bm{e_t}$ is the target people's voiceprint and $\bm{\hat{e}}$ is the voiceprint of different types of audio.}
    \label{fig:data_augmentation}
    \vspace{-7mm}
\end{figure}

\begin{table}[H]
    \vspace{-3mm}
    \caption{Phonetical properties for data augmentation}
    \centering
    \resizebox{\columnwidth}{!}{
        \begin{tabular}{cccc}
        \toprule
        \multirow{2}{*}{\begin{tabular}[c]{@{}c@{}}Phonetical\\ Properties\end{tabular}} & \multirow{2}{*}{\begin{tabular}[c]{@{}c@{}}Modification \\ Range\end{tabular}} & \multicolumn{2}{c}{Emotional Impact}   \\ \cline{3-4}& & $\uparrow$ & $\downarrow$ \\ 
        \midrule
        Speech Rate     & 0.3-1.8 & Fear or Disgust        &  Sadness   \\
        F0 Mean      & 0.9-1.1 &  Anger or Happiness    & Disgust or Sadness  \\
        F0 Contour        & 0.7-1.3 & Anger or Happiness & Sadness \\
        Energy          & 0.5-2.0 & -    & -     \\ 
        Sequential Order    &      -  & -    & -     \\  
        \bottomrule

        \end{tabular}
    }
    \label{tab:phonetical_properties}
    \vspace{-3mm}
\end{table}

\textit{Speech rate:} We change the original speech rate with a factor $\alpha$ using FFmpeg~\cite{ffmpeg} and $\alpha$ is uniformly sampled from $[0.3,1.8]$, which is obtained experimentally to guarantee an acceptable impact on phonetical properties. Please note that the change of speech rate here is independent of the acceleration of $S_1$ in the noise design, which aims to increase the number of vowels per unit time of our noise.

\textit{F0 mean:} Similar as before, we vary the F0 mean by a multiplicative factor $\alpha$ which is uniformly sampled from $[0.9,1.1]$. This range is also determined experimentally to limit the impact on phonetical properties.

\textit{F0 contour:} It shows the audio's pitch along with time. We exaggerate or flatten the F0 contour with $F0^{'} = \alpha(F0 - \overline{F0}) + \overline{F0}$ using World vocoder~\cite{world_vocoder}. The factor $\alpha$ is uniformly sampled from $[0.7, 1.3]$.

\textit{Energy:} We modify the average amplitude of the data with a factor randomly sampled from $[0.5, 2]$.

\textit{Sequential order:} For human speech data, the reverse in the time domain has little effect on phonetical properties but will disrupt the semantic information. In this paper, we randomly reverse each phoneme data in the time domain.

To visualize the impact of data augmentation, we extract voiceprints from each audio and calculate the cosine distance between them and the average voiceprint of the target people.
We compare four types of data: the data from the same people, from the people who have the closest voiceprint to the target people among the dataset (LibriSpeech train-clean-100~\cite{LibriSpeech}), the data processed by all five augmentation methods, and processed by every single method. Results in Figure~\ref{fig:data_augmentation} show that augmenting the data with a single method has limited impacts on its phonetical features. Even when augmented by all the five methods, the data is more similar to the original data than that of the closest people in the dataset. 

\textbf{Noise Generation.} With the augmented phoneme data, we now can generate noise with the method stated in Section~\ref{subsec:noise_design}. It should be noted that different from the origin method, the phoneme data used in this process could be either recorded from the target people, or matched from a pre-prepared dataset as stated in Section~\ref{subsec:user_registration}.

\begin{figure}[t]
    \centering
    \begin{minipage}[t]{0.48\linewidth}
        \centering
        \includegraphics[width=\linewidth]{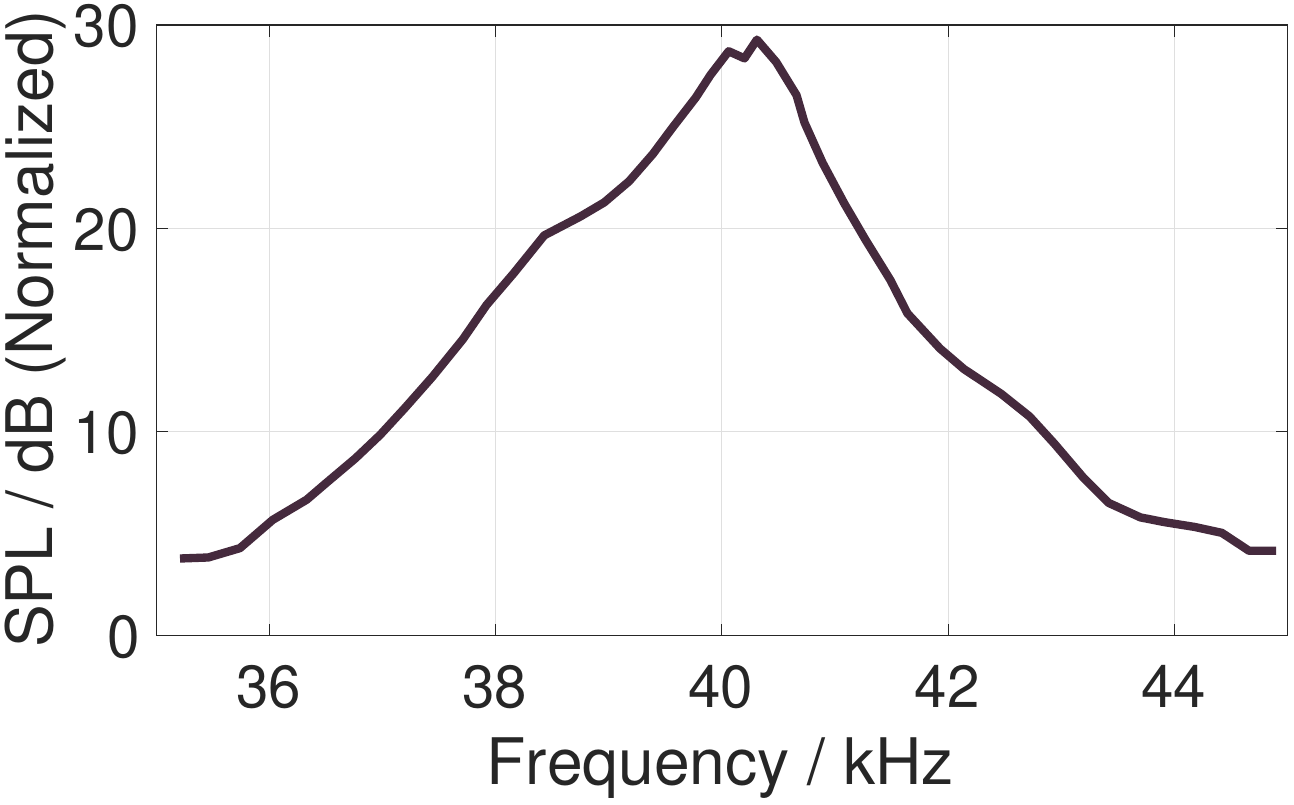}
        \caption{Frequency response of the transmitter.}
        \label{fig:transmitter_fr.pdf}
    \end{minipage}
    \begin{minipage}[t]{0.48\linewidth}
        \centering
        \includegraphics[width=\linewidth]{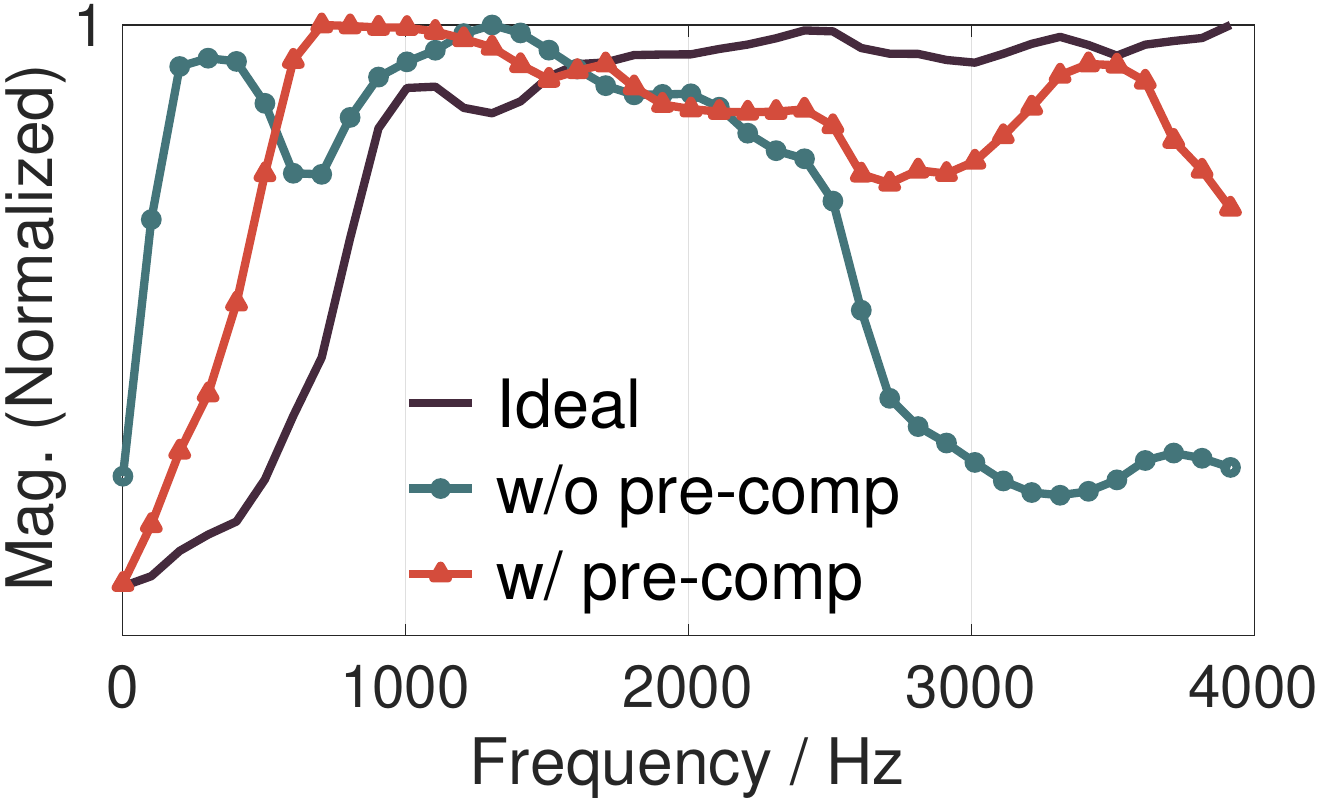}
        \caption{Spectrum of recorded sweep signals.}
        \label{fig:w_o_pre_compensation}
    \end{minipage}
    \vspace{-5mm}
\end{figure}

\textbf{Pre-Compensation.} Before we transmit the generated noise with ultrasound transmitters, it is necessary to compensate the noise signal due to the non-flat frequency response (FR) of transmitters. Figure~\ref{fig:transmitter_fr.pdf} illustrates a typical FR of a ultrasound transmitter~\cite{jinci} with a central frequency of 40 kHz, which exhibits a peak at about 40 kHz and rapidly attenuates as the frequency moves away. When transmitting noises with a carrier of 40 kHz (amplitude modulation), the low-frequency part will be promoted while the high-frequency part will be suppressed, as shown in Figure~\ref{fig:w_o_pre_compensation}, which could cause distortion of the noise thus decreasing the jamming effectiveness.

To address this, we analyze the FR of the recorder: $H_2(f)$, and the equivalent FR between the transmitter and the recorder: $H_1(f)$. The FR of the transmitter could be roughly estimated as $ H_1(f) * \frac{1}{H_2(f)}$. Then an inverse filter $h_1^{-1}(t) \ast h_2(t)$ \footnote[1]{$h(t) \xleftrightarrow {\mathcal{F}} H(f)$ means Fourier transform and $\ast$ means convolution.} is applied on the noise signal before modulation to compensate the unwanted distortion. We analyze $H_1(f)$ and $H_2(f)$ using multiple recorders and then adopt the average. Figure \ref{fig:w_o_pre_compensation} shows that the compensation works evidently. Please note that only the frequency blow 4kHz is considered for compensation as the majority of energy in human voice falls in this range. The drop below 1kHz in the ideal spectrum is caused by the recorders' imperfection.

\textbf{Low-Sideband Noise Modulation.} 
With the compensated jamming noise, we now proceed to modulate it onto a carrier and transmit it via ultrasound transmitters to jamming microphones stealthily. However, the two widely used modulation methods, double sideband amplitude modulation (DSB-AM) \cite{10.1145/3133956.3134052,patronus} and frequency modulation (FM) \cite{backdoor}, are not suitable here. The wide frequency range of our noise results in a large bandwidth of the modulated signal if DSB-AM is applied, which increases the audibility of the noise because of the self-demodulation in the transmitter~\cite{backdoor, 211283}. As for FM, our phoneme-based noise cannot be recovered by self-demodulation at the receiver end.

In this paper, we use single-sideband modulation (SSB-AM) to reduce the noise audibility. Take the lower-sideband amplitude modulation (LSB-AM) as an example, assume the carrier signal is $cos(2\pi f_c t)$ and the noise signal is $n(t)$, then the DSB-AM can be represented by $n_D(t) = \sqrt{2}n(t) cos(2\pi f_c t)$ and the LSB-AM can be represented by $n_L(t) = n(t) cos(2\pi f_c t) + \widehat{n}(t) sin(2\pi f_c t)$, where $\hat{n}(t)$ is the Hilbert transform of $n(t)$ and $f_c$ is the carrier frequency. The coefficient $\sqrt{2}$ in the former is to ensure the energy of the two modulated signals are the same.
 
Similar to microphones, the nonlinearity in the ultrasonic transmitter also generates high order components. Because energy decays significantly as the order goes up, we only consider the quadratic term here. With DSB-AM and LSB-AM modulation, the quadratic terms of the modulated signal can be represented by Equations \ref{equ:DSB_qua} and \ref{equ:LSB_qua}. As shown in these two equations, the human audible low-frequency components for DSB-AM is $n^2(t)$, while for LSB-AM it is $\frac{1}{2}(n^2(t) + \widehat{n}^2(t))$. Because the Hilbert transform imparts a phase shift of $\pm \frac{\pi}{2}$ to each frequency component, the amplitude of each frequency component in the former is $\sqrt{2}$ times bigger than the latter. That is, the former has twice the energy of the latter which means SSB-AM can reduce the power of the audible sound generated at the transmitter to 0.5 times the original. 

\begin{equation}
    \begin{aligned}
     n^2_{D}(t) = &n^2(t)(1+cos(4\pi f_c t))\xRightarrow[Filter]{Lowpass} n^2(t)
    \end{aligned}
    \label{equ:DSB_qua}
\end{equation}
\begin{equation}
    \begin{aligned}
        n^2_{L}(t) \xRightarrow[Filter]{Lowpass}0.5(n^2(t) + \widehat{n}^2(t))
    \end{aligned}
    \label{equ:LSB_qua}
\end{equation}

Additionally, we choose lower-sideband instead of higher-sideband because of its lower attenuation in the air. The $f_c$ we used is 40 kHz, which has the maximum resonance response in most microphones~\cite{backdoor}.

We conduct an experiment with 3 males and 3 females aged from 22 to 31~\footnote[1]{All procedures performed in studies involving human in this paper are validated through an institutional review (IRB)} to test the impact of modulation methods on audibility.
We gradually increase the transmission power until volunteers perceive the noises. The normalized maximum transmission energy is shown in Table~\ref{tab:self_demodulation}.

\begin{figure}[t]
    \centering
    \includegraphics[width=\columnwidth]{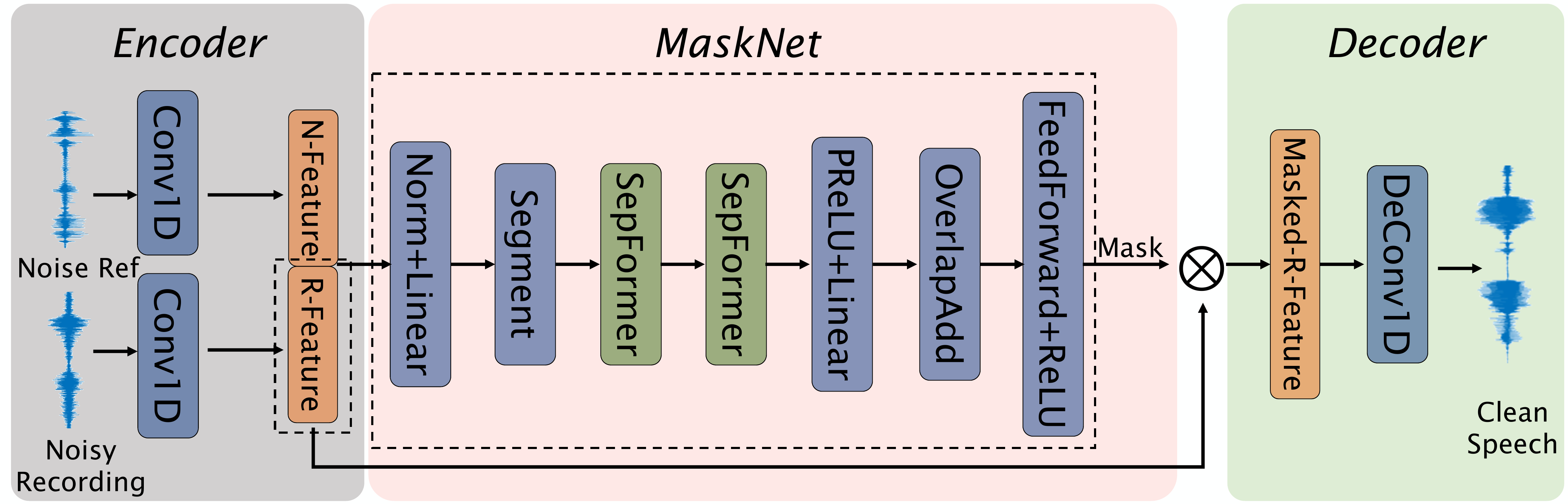}
    \caption{Network Architecture}
    \label{fig:network_architecture}
\end{figure}

\begin{table}[t]
    \centering
    \resizebox{0.9\columnwidth}{!}{
    \begin{tabular}[\textwidth]{cccc}
        \toprule
        \multirow{2}{*}{Noise} & \multicolumn{3}{c}{Normalized Energy} \\ \cline{2-4} 
                               & DSB-AM      & LSB-AM     & USB-AM     \\ 
        \midrule
        White Noise            & 1.00        & 1.49       & 1.29       \\
        Phoneme-Based Noise    & 2.77        & \textbf{4.14}       & 3.61       \\ 
        \bottomrule
    \end{tabular}}
    
    \caption{Normalized max transmission energy.}
    \label{tab:self_demodulation}
    \vspace{-5mm}
\end{table}

\subsection{Content Recovery}
In certain situations, users may need to record the ongoing conversation while ensuring protection against adversaries. However, all microphones within the environment would be affected by InfoMasker, resulting in noisy recordings for both adversaries and users. 

To address this problem, we plan to eliminate the jamming noise from noisy recordings with the help of deep-learning model. Specifically, assume the ongoing speech is $s(t)$ and the transmitted jamming noise is $n(t)$, the noisy recording could be represented by:
$$
r(t) = s(t) + \alpha \cdot h_{1}(t) \ast n(t) + a(t)
$$
where $h_{1}(t)$ is the combination of channel impulse response (CIR) caused by over-the-air transmission and the residual distortion of the compensation process for the jamming noise. $a(t)$ represents the ambient noise. $\alpha$ is an unknown coefficient representing the amplification factor of the noise. 

Our target is to acquire $s(t)$. Our main idea is to remove noise components (i.e., $\alpha \cdot h_{1}(t)\ast n(t) + a(t)$) from noisy recordings $r(t)$. Because adversaries only have access to $r(t)$, it is hard for them to recover $s(t)$, while it is possible for the user with the help of the detail of the jamming noise signal $n(t)$. InfoMakser could store these noises along with timestamps while jamming recorders. This additional information makes it possible to recover $s(t)$.

\textbf{Network Architecture.} To achieve this target, we design a transformer-based denosing network based on~\cite{sepformer}. The network takes the noisy recording and the clean noise reference as inputs, generating clean audios as the output, as illustrated in Figure~\ref{fig:network_architecture}, which contains three parts: encoder, masknet, and decoder. The encoder employs two convolution layers to convert input audios to 2-D feature maps. Then the feature maps of the noisy recording and the noise reference are concatenated and fed into the masknet. The masknet contains two SepFormer blocks that can focus on both short- and long-term audio features, enabling it to find connections between these two feature maps and generate a mask to filter out noise components from the noisy recording's feature map. The decoder is symmetrical to the encoder. containing deconvolution layer and converting masked audio feature maps to clean audios.

We adopt the scale-invariant signal-to-noise ratio (SI-SNR)~\cite{si-snr} as the loss function, as shown in Equation~\ref{equ:sisnr}. Compared with the SNR metric, SI-SNR is more robust, scale-invariant, and can characterize the speech quality.

\begin{equation}
    \mathcal{L}_{\text{SI-SNR}}(s, \hat{s}) = -10 \text{log}_{10} \left( \frac{||\frac{\hat{s}^T s}{||s||^2}s||^2}{||\frac{\hat{s}^T s}{||s||^2}s - \hat{s}||^2} \right)
    \label{equ:sisnr}
\end{equation}

\textbf{Training Process.} There are two main challenges recovering the clean audio: \textbf{\textit{Time-domain misalignment}}: There are two reasons account for this. The first is that it takes different time for the voice and the jamming noise to arrive at the recorder. For example, the difference of 1 meter in transmission distance can make the error between two timestamps up to $1/340\approx0.0029$ seconds, about 139 sampling points (48 kHz). And the second is that the timestamps of audio recordings are often coarse.
For example, the recording timestamp in iOS 16.1 (iPhone 14) and Android 12 (Redmi K30Ultra with MIUI 14.0.1) is often accurate to second, which means the error could be up to 48000 sampling points under a 48 kHz sampling rate.
\textbf{\textit{Unknown channel impulse responses}}: The usage environment of InfoMasker could be various, which means the channel impulse responses of the target speech signal and the jamming noise are uncertain and could differ significantly from each other. So we can not preset $h_{1}(t)$ in the training process.

To cover the first challenge, we introduce a random shift in the noise reference in the training process, and the shift interval is [-fs, fs] where fs is the audio sample rate. The random shift will force the network to align the noisy recording and the reference automatically. While for the second, we leverage several public acoustic CIR datasets, including~\cite{AIR, RWCP, SPARG}, and randomly choose $h_{1}(t)$ from them for each training step. The abundant CIR data in the training process could help the neural network learn characteristics of reverberation and eliminate the effect of reverberation when generating the mask.

\section{Hardware Design}
\label{sec:hardware_design}

\subsection{Characteristics of Transmitters}

\begin{figure}[t]
    \centering
    \includegraphics[width=\linewidth]{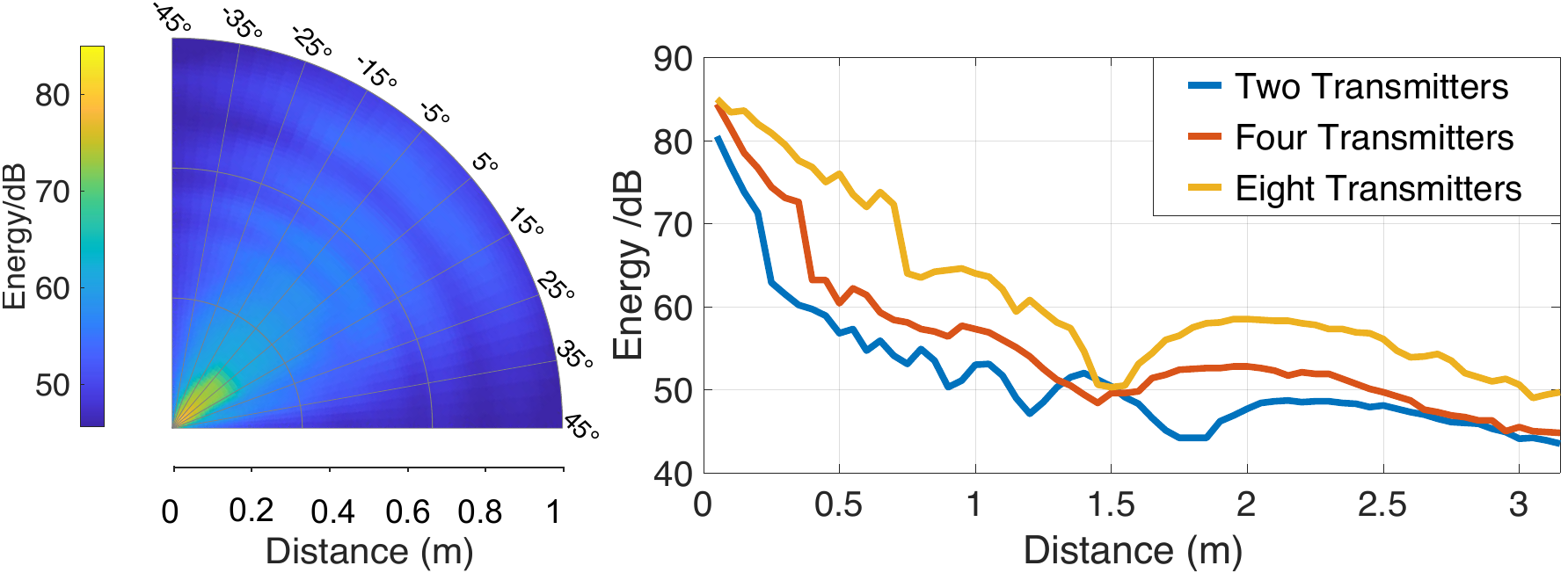}
    \caption{Characteristics of the transmitter. Left: Energy distribution of two transmitters. Right: Energy attenuation with distance when the angle is $0^\circ$.}
    \label{fig:energy_distribution}
    \vspace{-5mm}
\end{figure}

We first study the characteristics of a widely-used off-the-shelf transmitter~\cite{jinci}
to prepare for our array design. To simulate the noise injection scenario, we play 39kHz and 41kHz single tones through two transmitters separately and measure the energy of the demodulated signal around transmitters. The result in the left of Figure~\ref{fig:energy_distribution} shows that the energy around transmitters attenuates rapidly with angle, which drops nearly 10 dB within $25^\circ$ and so cannot meet the jamming requirements in the real-world. There are two reasons account for this. First, ultrasound propagates more straight compared to human audible sound. Second, the success of noise injection requires both the carrier and the modulated noise signal arrive at the recorder. We further analyze the energy attenuation with varying distances when different number of transmitters are deployed. As shown in the right of Figure~\ref{fig:energy_distribution}, although the energy attenuates rapidly as distance increases, we also witness the increase of ultrasound energy via utilizing more transmitters.

The results provide two valuable insights: 1. We can increase the effective jamming distance by simply increasing the number of transmitters. 2. To increase the effective jamming angle, it is necessary to further design the distributions of the transmitters for both carrier and modulated noise.

\subsection{Transmitter Array Design}
\begin{figure*}
    \centering
    \begin{minipage}[t]{0.75\textwidth}
        \begin{subfigure}[t]{0.51\linewidth}
            \begin{subfigure}[b]{0.3\linewidth}
                \centering
                \includegraphics[width=\linewidth]{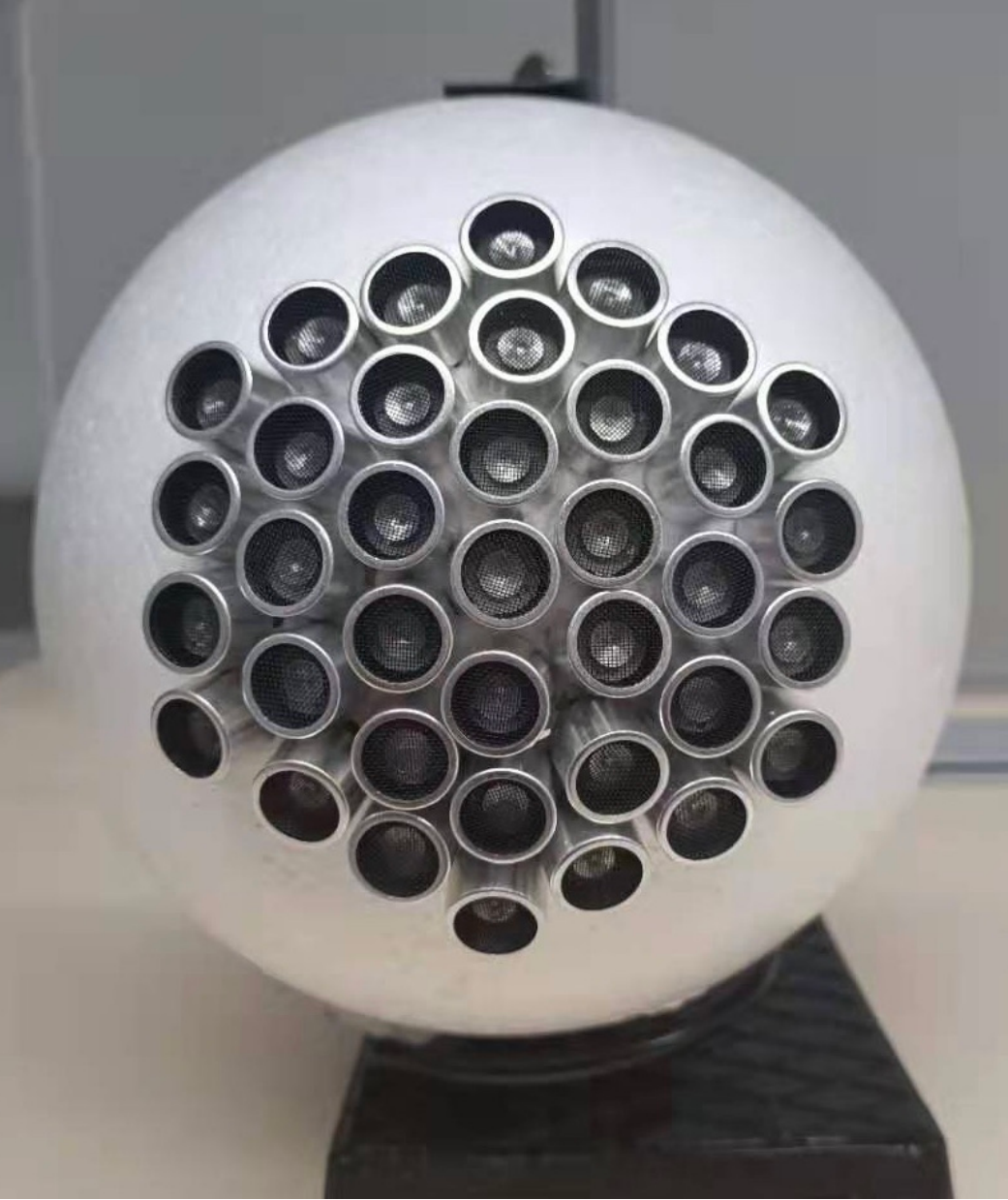}
            \end{subfigure}
            \begin{subfigure}[b]{0.3\linewidth}  
                \centering 
                \includegraphics[width=\linewidth]{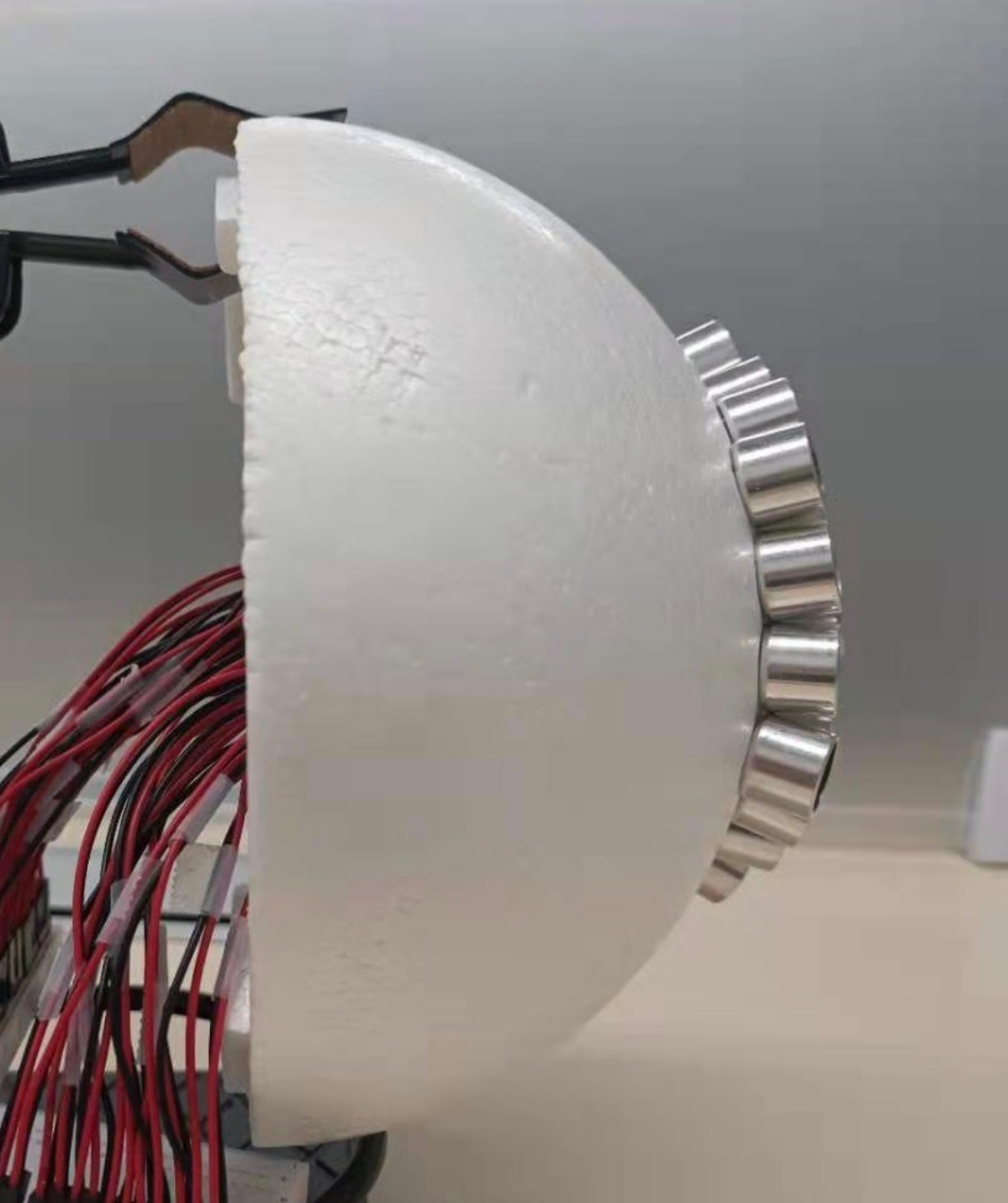}
            \end{subfigure}
            \begin{subfigure}[b]{0.3\linewidth}   
                \centering 
                \includegraphics[width=\linewidth]{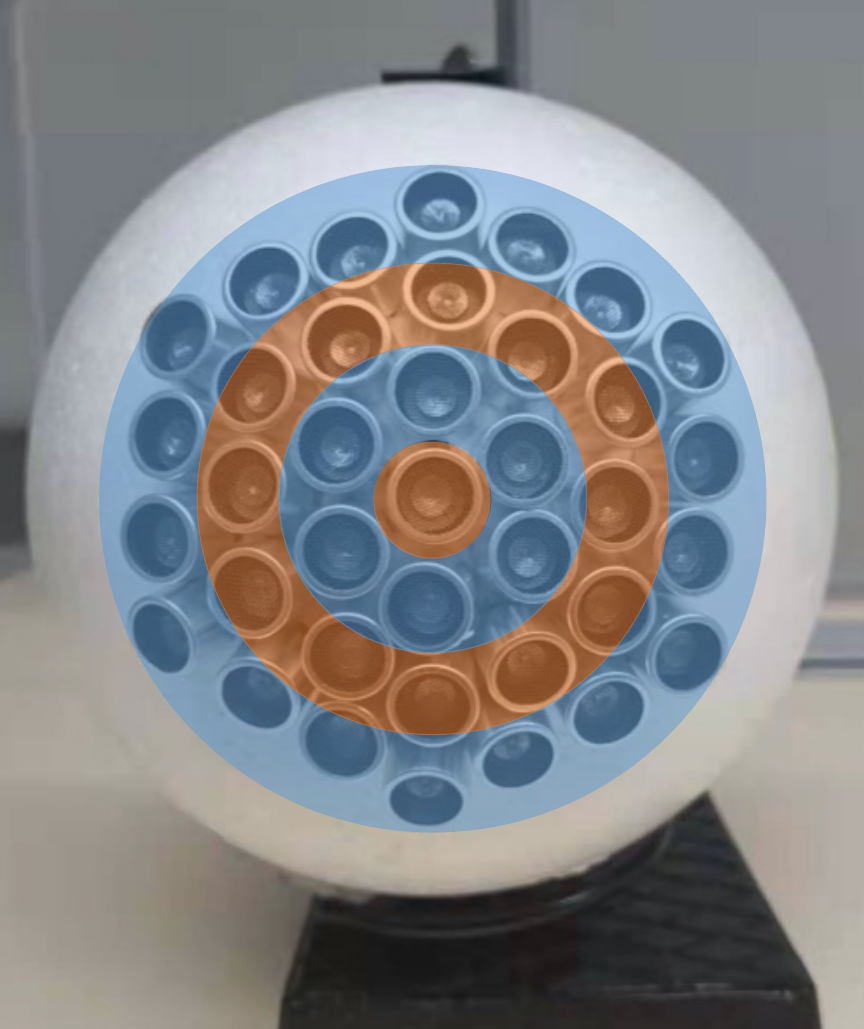}
            \end{subfigure}
        \end{subfigure}
        \begin{subfigure}[t]{0.44\linewidth}
            \centering
            \includegraphics[width=\linewidth]{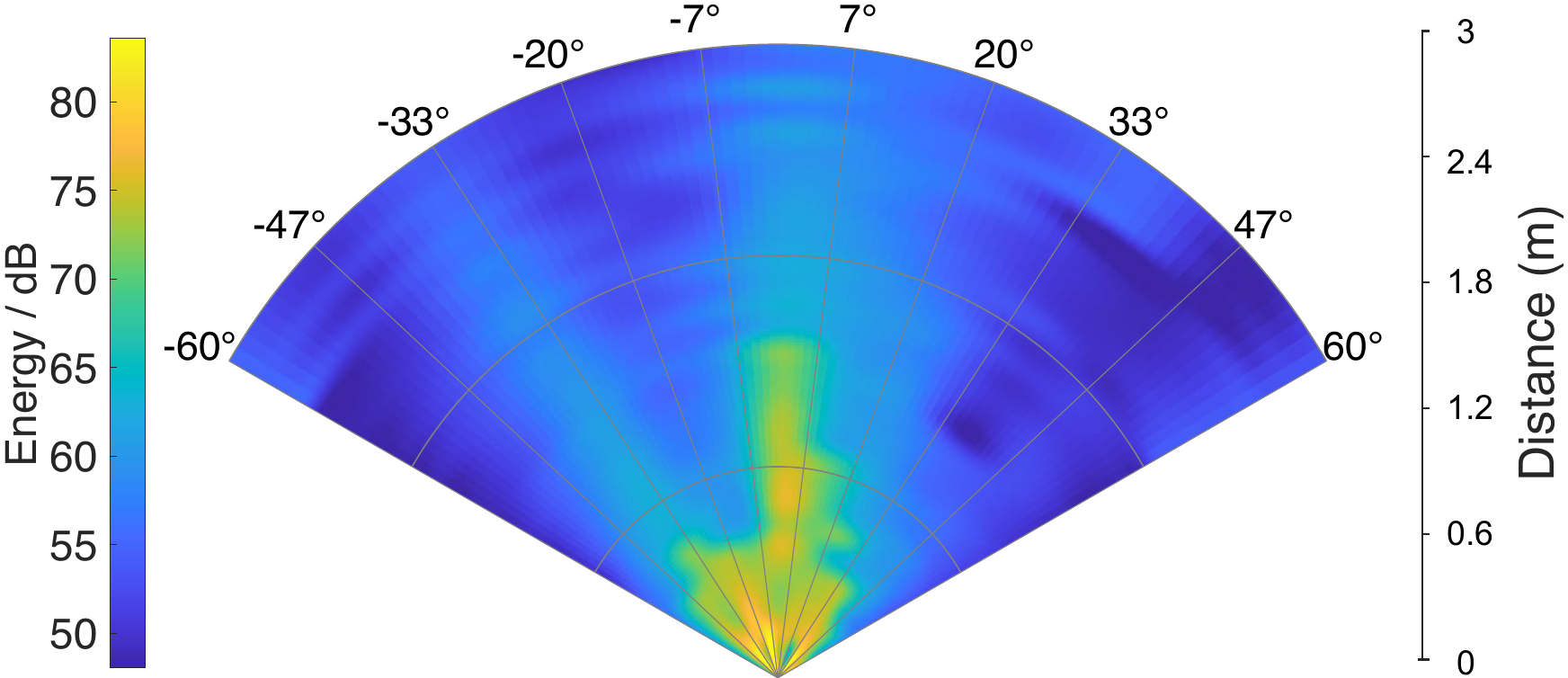}
        \end{subfigure}
        \caption{The transmitter array. Left: The orange transmitters transmit the carrier signal while the blue ones transmit the modulated noise signal. Right: Energy distribution of the array}
        \label{fig:transmitter_array}
    \end{minipage}
    \begin{minipage}[t]{0.23\textwidth}
        \centering
        \includegraphics[width=\linewidth]{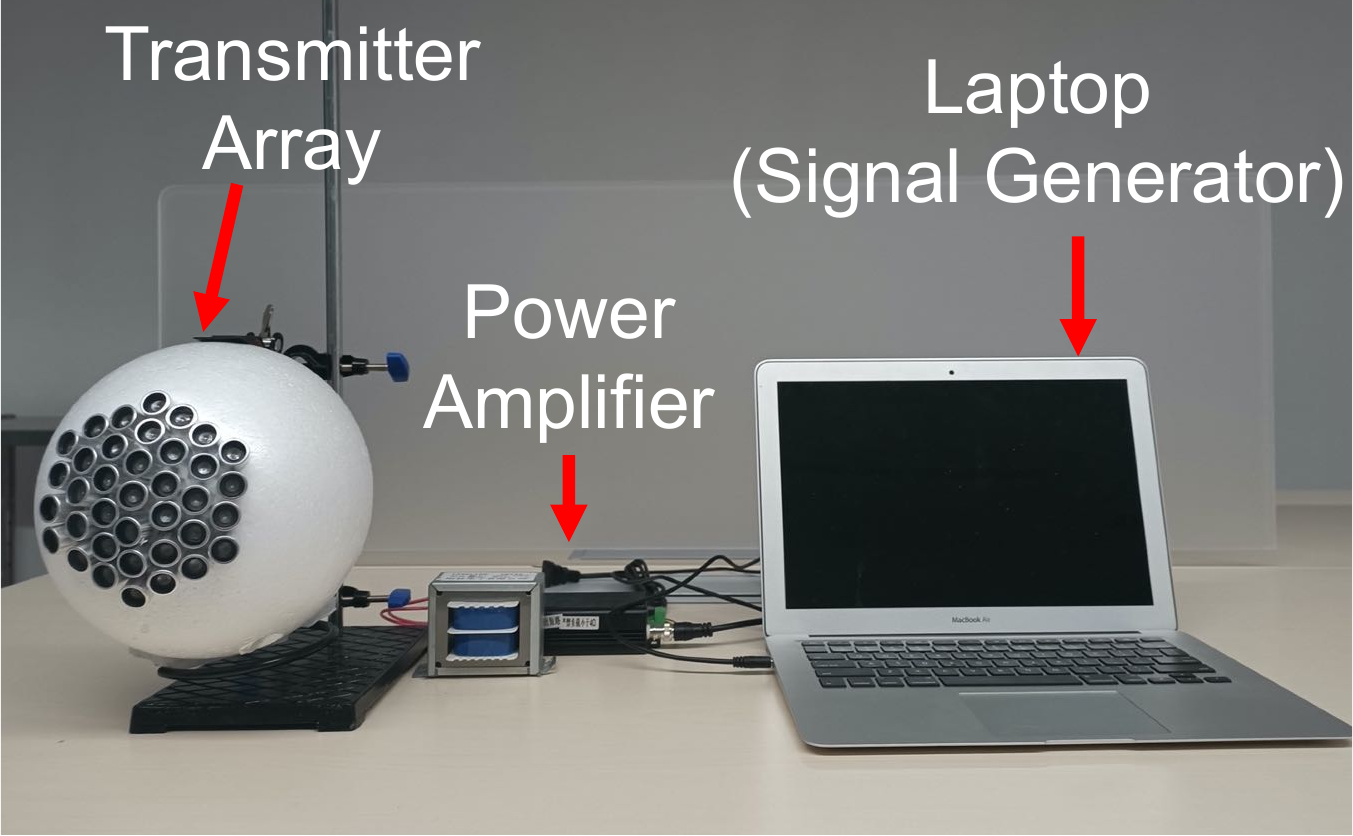}
        \caption{Hardware implementation.}
        \label{fig:hardware_implementation}
    \end{minipage}
    \vspace{-5mm}
\end{figure*}

 To extend the coverage of the transmitter, we design a transmitter array as shown in Figure~\ref{fig:transmitter_array}~(left). We install transmitters in a hexagonal shape on the spherical foam and separate them into two groups: one group sends the carrier signal and the other sends the modulated noise signal. The curvature of the spherical foam and the distribution of the transmitters enable the large coverage of the transmitter array. The energy distribution around the transmitter array is shown in Figure~\ref{fig:transmitter_array}~(right) and it is obvious that it can cover a large span of angle up to about $90^{\circ}$ and a much longer distance compared to using a single transmitter.

\subsection{Hardware Implementation}
The implementation is shown in Figure~\ref{fig:hardware_implementation}, including a transmitter array, two power amplifiers (one for each transmitter group), a signal generator ( We use a laptop here but it could be replaced by other development boards with a sound card sampling rate $\geq$ 80kHz). The content recovery is realized on a backend server (Intel(R) Xeon(R) Gold 6226R CPU and a Nvidia RTX3090 GPU in this paper). Without considering the laptop and the backend server, the hardware implementation costs about 70 dollars. 
\section{Evaluation}
\label{sec:eval}

\subsection{Experimental Setup}
\textbf{Evaluation Methodology.}
In this part, we first systematically evaluate the effectiveness of out system, then provide a detailed analysis to evaluate its robustness against attackers with different capabilities. We also test our system under various variables. At last, a case study is conducted in a common office to validate the our system's practicality.

Please note that the inconsistency in the accuracy of Tencent ASR across experiments is caused by differences in test sets and test dates. But for each experiment, the test set is consistent and is done within a short time period.

\textbf{Dataset.}
We adopt 6 widely-used datasets in different parts of this section. LibriSpeech~\cite{LibriSpeech} is used in the baseline and other unmentioned parts. AISHELL-1~\cite{aishell_2017} (Mandarin), Multilingual LibriSpeech~\cite{mls2020_interspeech} (Portuguese), and Japanese Versatile Speech~\cite{takamichi2019jvs} are used for testing different languages.
TIMIT~\cite{timit} is used in Section~\ref{subsec:robustness_adversarial_training} because of its small size which can make machine learning models converge faster. Harvard Sentences~\cite{harvard_sentences} is used for human perception test because of its short sentences, which can reduce the difficulty for human to recognize.

\textbf{Dataset Preprocessing.}
Before generating noise, we need to first segment the corpus into phonemes. We segment the English corpus (LibriSpeech and Harvard Sentences) with Prosodylab Aligner~\cite{prosodylab_aligner} and Mandarin corpus with Charsiu~\cite{zhu2022charsiu}. For Portuguese and Japanese, we adopt the Montreal Forced Aligner~\cite{MFA}. The TIMIT dataset provides aligned phonemes initially. We separate the phonemes into vowels and consonants after segmentation according to the International Phonetic Alphabet (IPA).

\subsection{Effectiveness}
\label{subsec:baseline}

\textbf{Overall Performance.}
We first evaluate our noise in the single-user scenario. We test the jamming performance of our noise under different ASRs and use a [0, 8] kHz band-limited gaussian white noise for comparison. A test set containing 27000 words is generated from LibriSpeech.
Since built-in noise reduction mechanism of recording devices may affect the reception of white noise (proven in Fig~\ref{fig:white_denoise}), we directly mix noises with speech signal in digital domain and feed the mixed signals to ASRs for a fair comparison.

\begin{figure*}[t]
    \centering
    \includegraphics[width=1\linewidth]{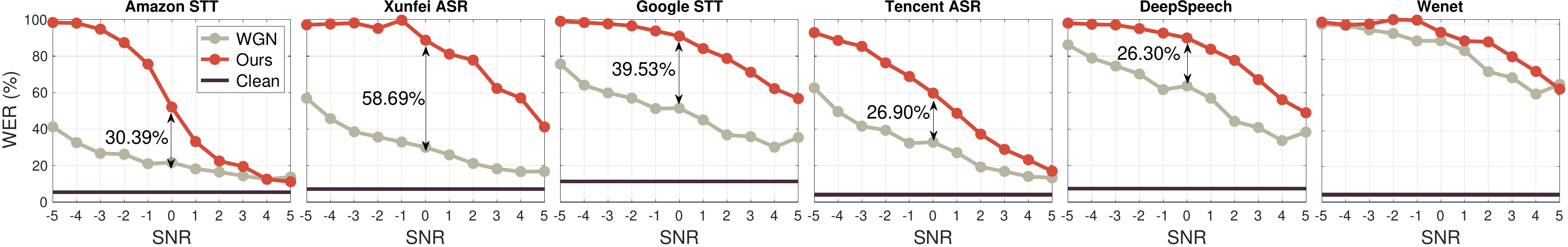}
    \caption{Recognition results of different ASR systems. We use the clear audio as a baseline.}
    \label{fig:digital_result_different_SNR}
    \vspace{-5mm}
\end{figure*}

We test four commercial ASR systems (Amazon Transcribe \cite{aws}, Tencent ASR \cite{tencent_asr}, Xunfei ASR \cite{xunfei_asr} and Google Speech-to-Text \cite{google_stt}) and two commonly used open source ASR systems (DeepSpeech \cite{deepspeech} and WeNet \cite{yao2021wenet}). The results in Figure \ref{fig:digital_result_different_SNR} show that our noise performs significantly better than white noise when $SNR \leq 4$, and the gap between them gradually increases as the SNR decreases. Besides, the advantage of our noise is more obvious on commercial ASRs than on open-source ones. We suppose this is because commercial systems have been enhanced for the interference of white noise.

\begin{figure}[t]
    \centering
    \includegraphics[width=\linewidth]{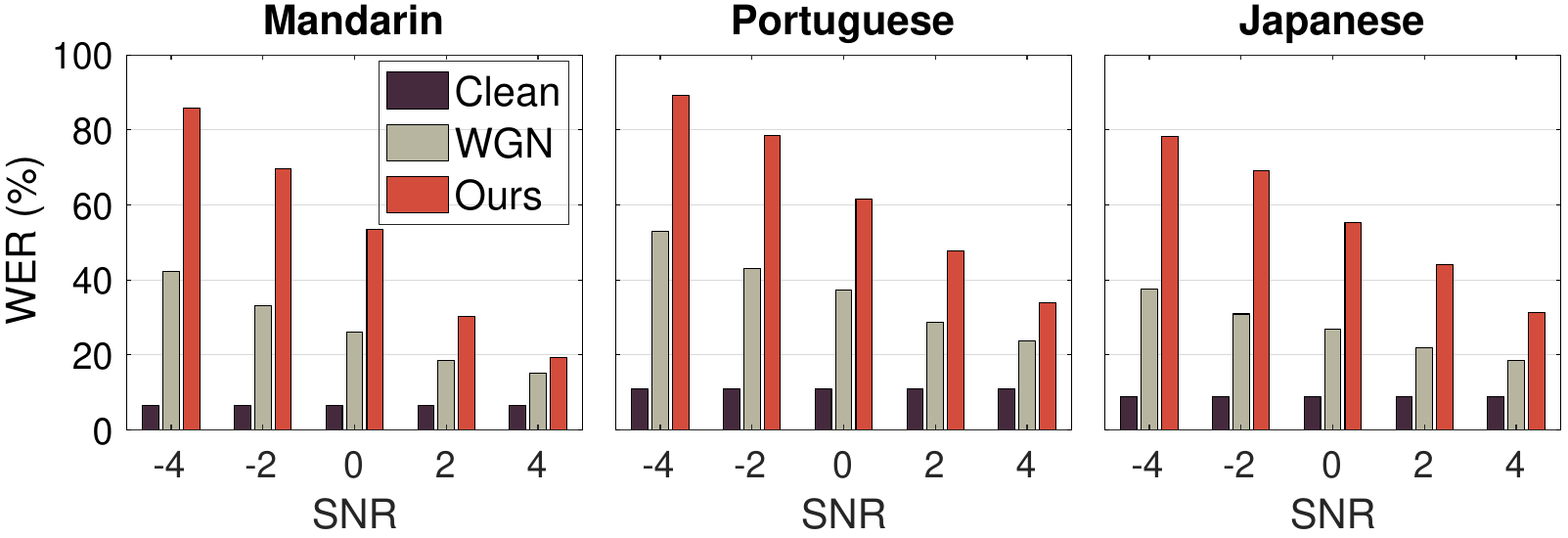}
    \caption{Results when applying on different languages. White gaussian noise (WGN) is also tested for comparisons.}
    \label{fig:multi_language}
\end{figure}

\textbf{Different Languages.}
We choose other three languages from different language families to ensure sufficient phonemic diversity. In addition to English (West Germanic language in Indo-European language family), we test Mandarin (Sino-Tibetan language family), Portuguese (Western Romance language in Indo-European language family) and Japanese (Japonic language family). We randomly choose 200 audios from each dataset as the test set and mix them with corresponding noises under different SNRs. Specifically, we recognize the Portuguese data with Amazon Transcribe because of its good performance on clean Portuguese corpus. The result in Figure~\ref{fig:multi_language} shows that our jamming method performs well in different languages.

\begin{figure}[t]
    \centering
    \includegraphics[width=\columnwidth]{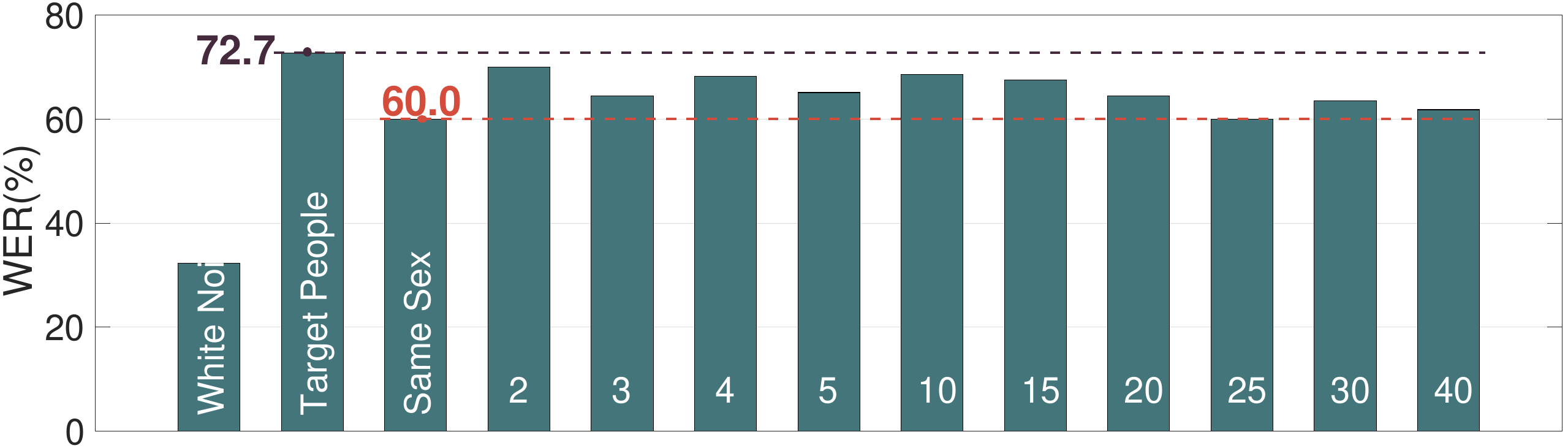}
    \caption{Recognition results for multi-user scenario. The numbers in the bar represent the number of user.}
    \label{fig:multi_user_result}
    \vspace{-5mm}
\end{figure}

\textbf{Multi-User Scenario.}
We evaluate our system with different number of randomly chosen users.
The results in Figure~\ref{fig:multi_user_result} shows that when the number of users is smaller than 10, the jamming effectiveness is lower than that of the single user scenario where the noise is generated based on the target person's speech, but significantly higher than the noise generated from the speech data of the other person with the same sex. When the number of users is greater than 10, the noise's jamming effectiveness gradually approaches to the noise generated from the people with the same sex.

\textbf{Real-World Scenario.} We place a smartphone around the transmitter array and adjust the transmitter's energy to get different SNRs. We use another smartphone to play speech signals. As it is hard to get a stable and precise SNR in real-world scenario, we test several times in each SNR interval and then calculate the average and the minimum WER. We collect more than 70 hours data in totally. The results in Table \ref{tab:real_world_wer} show that when $SNR < 0$, the WER in real-world scenario is slightly lower than that in the digital domain, but significantly higher when $SNR > 0$. 

\begin{table}[h]
    \vspace{-3mm}
    \caption{Recognition result in real-world scenario.}
    \centering
    \resizebox{\columnwidth}{!}{
        \begin{tabular}{cccccccc}
        \toprule
        \textbf{SNR}         & \textbf{\textless{}-4} & \textbf{{[}-4,-2{]}} &\textbf{ {[}-2, 0{]} }&\textbf{ {[}0,2{]}} &\textbf{{[}2,4{]}} & \textbf{\textgreater{}4 }&\textbf{ Clear} \\ 
        \midrule
        \textbf{Avg WER(\%)} & 85.8          & 81.6        & 77.6        & 70.2      & 56.4      & 42.3            & 11.5  \\ 
        \textbf{Min WER(\%)} & 68.6          & 77.0        & 62.4        & 62.2      & 45.3      & 30.3            & -     \\ \hline \hline
        \textbf{Digital WER(\%)} & 88.6          & 85.4       & 68.8        & 48.67      & 28.9      & 17.0           & 4.1     \\ 
        \bottomrule
        \end{tabular}
    }
    \label{tab:real_world_wer}
    \vspace{-3mm}
\end{table}

\begin{figure}[t]
    \centering
    \begin{subfigure}[t]{0.49\linewidth}
        \centering
        \includegraphics[width=\linewidth]{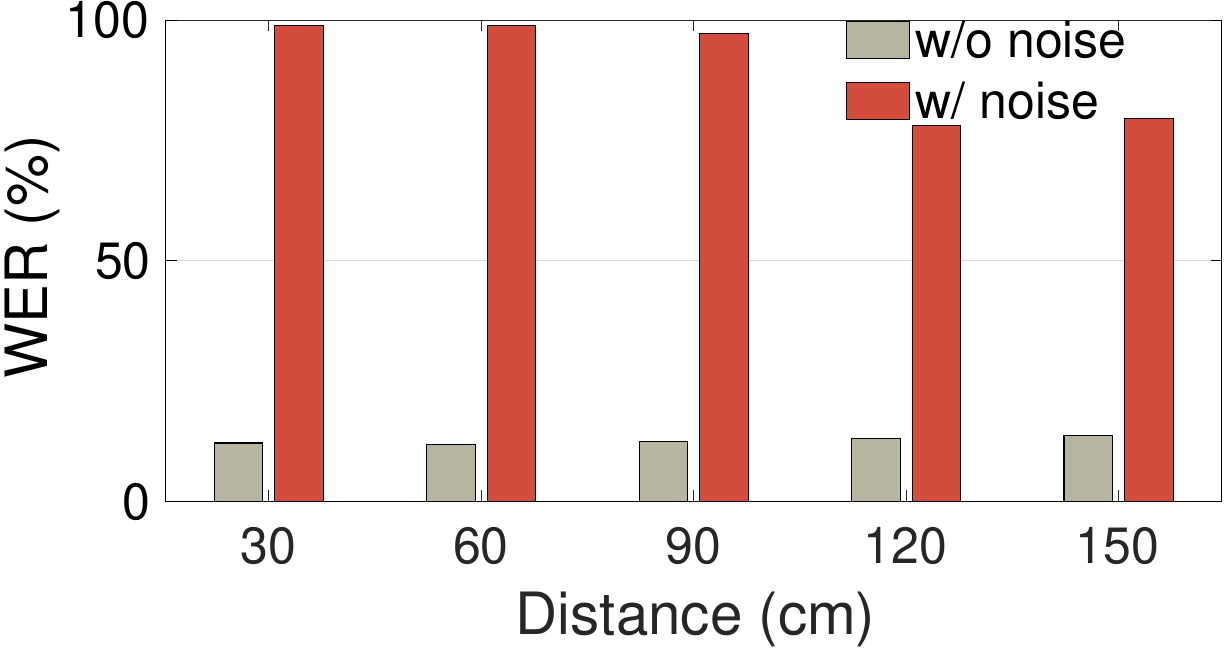}
    \end{subfigure}
    \begin{subfigure}[t]{0.49\linewidth}
        \centering
        \includegraphics[width=\linewidth]{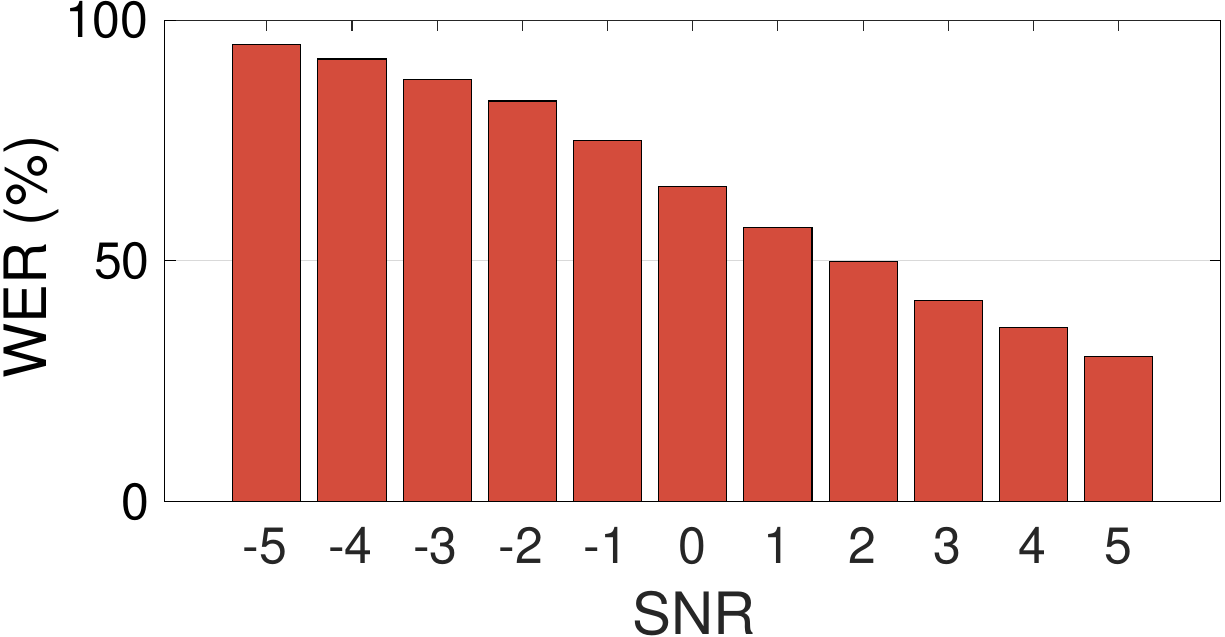}
    \end{subfigure}
    \caption{Results of the end-to-end scenario. Left: results in different distances. Right: results in different SNR.}
    \label{fig:end_to_end}
    \vspace{-5mm}
\end{figure}

\textbf{End-to-End Scenario.}
\label{subsubsec:real_world_end_to_end}
We further evaluate our system in a real-world end-to-end scenario with two volunteers (one male and one female). Each of them reads one sentence (about 5 seconds) randomly selected from~\cite{harvard_sentences} for registration. We extract their voice embeddings and then match the corresponding closest people in LibriSpeech for noise generation. To simulate a realistic scenario, we place the transmitter array on an office table and then place 5 smartphones acting as eavesdroppers with different distances to the array. We have the volunteer sit at the table and read 50 sentences selected from~\cite{harvard_sentences}. As it is hard to control the SNR precisely in a real-world scenario, we record the noise and speech separately and then mix them under various SNRs. The recognition results of the mixed audios (i.e., jammed speech) are shown in Figure~\ref{fig:end_to_end}.  The results show that our system performs well in the real-world end-to-end scenario.

\textbf{Human Perception.} We recruit 15 testees including 5 females and 10 males aged from 22 to 31 to test the recognizability of jammed audios. To make testees put their effort into the study, we adopt an accuracy-related reward to incent them. Tested audios are either generated in digital domain or recorded over-the-air. In the digital domain, in addition to white noise and our noise generated with the target's speech, we also test our noise generated from audios of same \& different sex people (SNR=-1). In the over the air setting, we test clean audios and jammed audios recorded in~\ref{subsec:case_study}. In addition to recognizing the audio, we also ask testees to score the recognizability of each audio from 0 to 5 where 5 indicates easily understandable and 0 means completely incomprehensible. In total, each testee needs to recognize 37 sentences selected from Harvard Sentences. The results in Table~\ref{tab:human_perception} show that our noise performs better than white noise. Besides, audios jammed by our noise also exhibit the worst recognizability.

\begin{table}[h]
    \vspace{-3mm}
    \caption{Human perception result. Audios of the right two types are recorded in the over-the-air scenario.}
    \centering
    \resizebox{\columnwidth}{!}{
        \begin{tabular}{cccccc||cc}
            \toprule
                & Clear & \begin{tabular}[c]{@{}c@{}}White\\ Noise\end{tabular} & \begin{tabular}[c]{@{}c@{}}Same\\ People\end{tabular} & \begin{tabular}[c]{@{}c@{}}Same\\ Sex\end{tabular} & \begin{tabular}[c]{@{}c@{}}Diff\\ Sex\end{tabular} & Clear & \begin{tabular}[c]{@{}c@{}}Same \\ People\end{tabular} \\ 
            \midrule

                WER (\%)    & 26.69  & 64.28  & 75.89  & 67.09  & 60.72  & 26.55  & 99.9  \\
                Avg. MOS (0-5) & 4.88     & 2.46 & 1.54 & 1.91 & 2.3 & 4.62  & 0.18 \\ 
            \bottomrule
        \end{tabular}}
    \label{tab:human_perception}
    \vspace{-7mm}
\end{table}

\subsection{Effectiveness of Content Recovery}
\label{subsec:effectiveness_of_content_recovery}
In this part we test the effectiveness of our recovery network. Additionally, we consider an adversary that tries to recover the recording without the noise reference. We adopt a similar network architecture shown in Figure~\ref{fig:network_architecture} to simulate the adversary. Specifically, we remove the noise reference part (the upper half of the encoder) and use only the noisy recording as the input. With the simulated adversary network, we process noisy recordings with four different methods: enhance~\cite{fullsubnetPlus} after denoised by our network or the simulated attacker network, only enhance, and no processing (origin noisy audios). We consider two scenarios: digital domain and real-world scenario, and randomly choose 200/50 audios as the testset from LibriSpeech test-clean for them, respectively. Similar to the real-world end-to-end scenario, here it is hard to control the SNR precisely in real-world. So we record the audios and the modulated noises separately and then mix them under various SNRs.

\begin{figure}[t]
    \centering
    \begin{subfigure}[t]{\linewidth}
        \includegraphics[width = \linewidth]{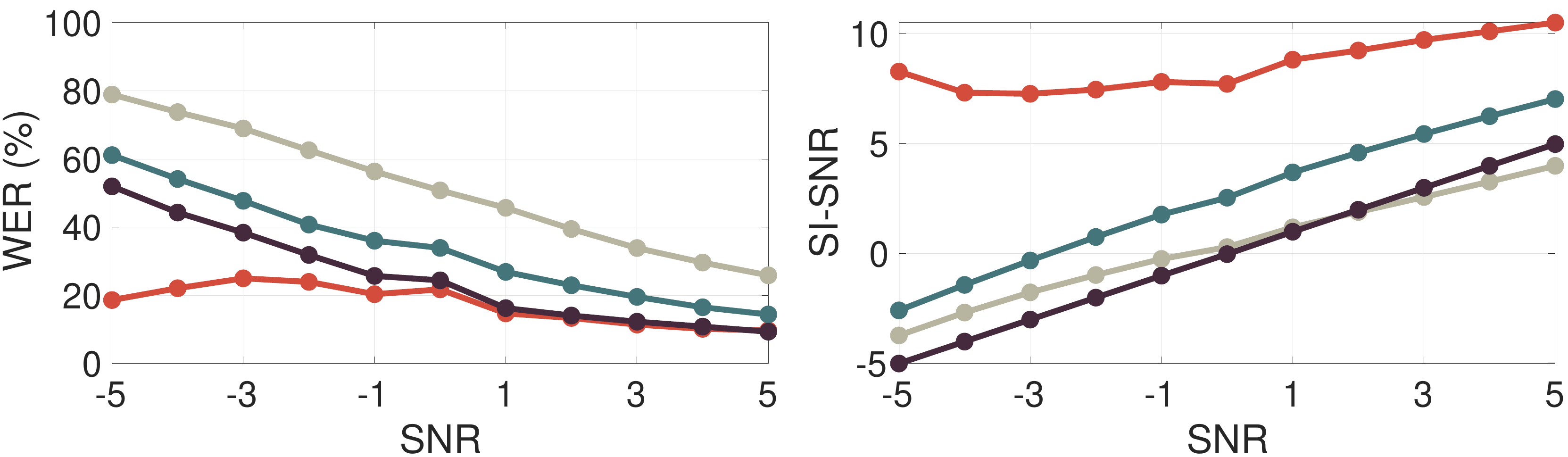}
    \end{subfigure}
    
    \begin{subfigure}[t]{\linewidth}
        \includegraphics[width = \linewidth]{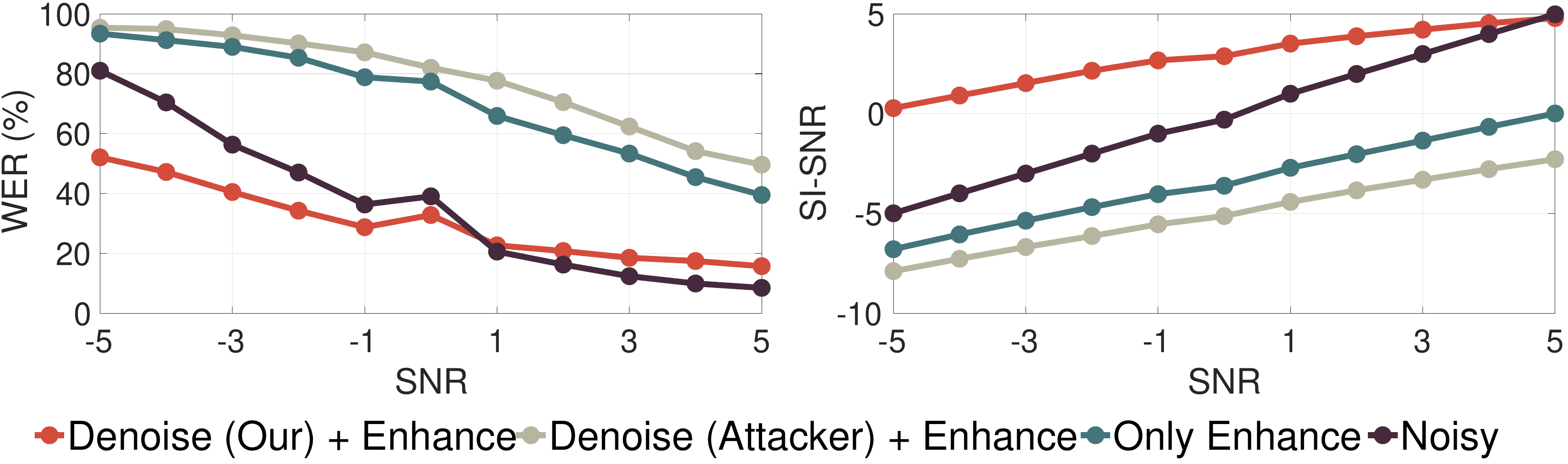}
    \end{subfigure}
    \caption{Result of content recovery. Top: digital scenario. Bottom: real-world scenario.}
    \label{fig:content_recovery}
    \vspace{-5mm}
\end{figure}

The results in Figure~\ref{fig:content_recovery} shows that when SNR$<$1, the content recovery network can improve the recognition accuracy significantly, with larger improvements at lower SNRs. When SNR$>=$1, the recognition accuracy drops slightly. While for the other two methods, enhance after denoising by simulated attacker network or only enhance, the recognition accuracy drops dramatically. Moreover, the audio quality (indicates by SI-SNR) is noticeably better after content recovery. These results show that our content recovery part can remove the jamming noise from the noisy recording, making it possible for the users to record the ongoing conversation.

\subsection{Robustness Against Various Attackers}
\label{subsec:robustness}
Here we evaluate the robustness of our system against attackers with different capabilities. We consider two types of attacker: the \textbf{normal attacker} who has no knowledge of our jamming method and the \textbf{advanced attacker} who knows the details of our jamming method. We conduct experiments under different attack scenarios and the results show that our system can resist both types of attackers effectively.

\begin{figure}[t]
    \centering
    \begin{minipage}[t]{0.49\linewidth}
        \includegraphics[width=\linewidth]{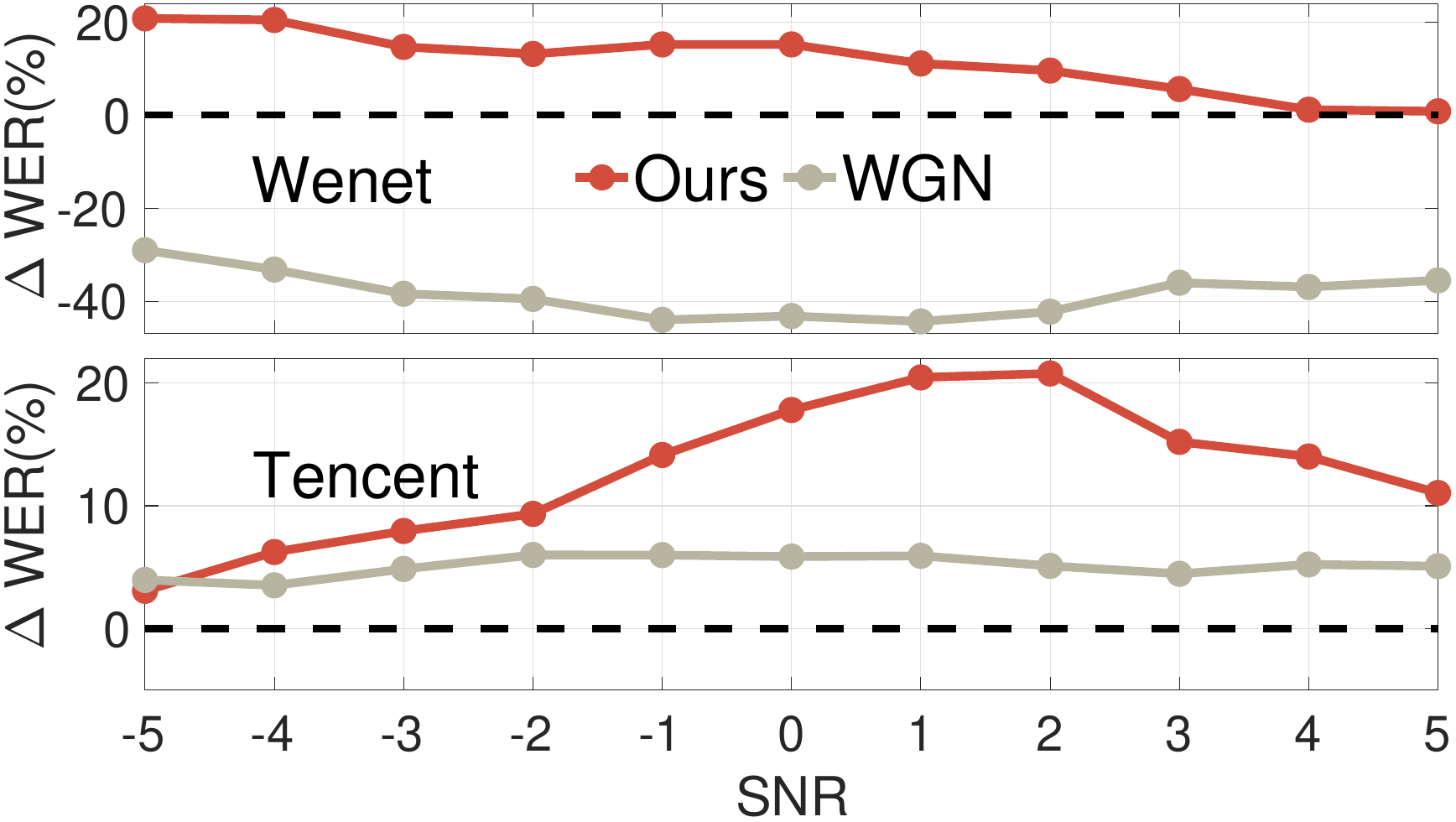}
        \caption{Recognition results after speech enhancement.}
        \label{fig:digital_enhanced}   
    \end{minipage}
    \begin{minipage}[t]{0.49\linewidth}
        \includegraphics[width=\linewidth]{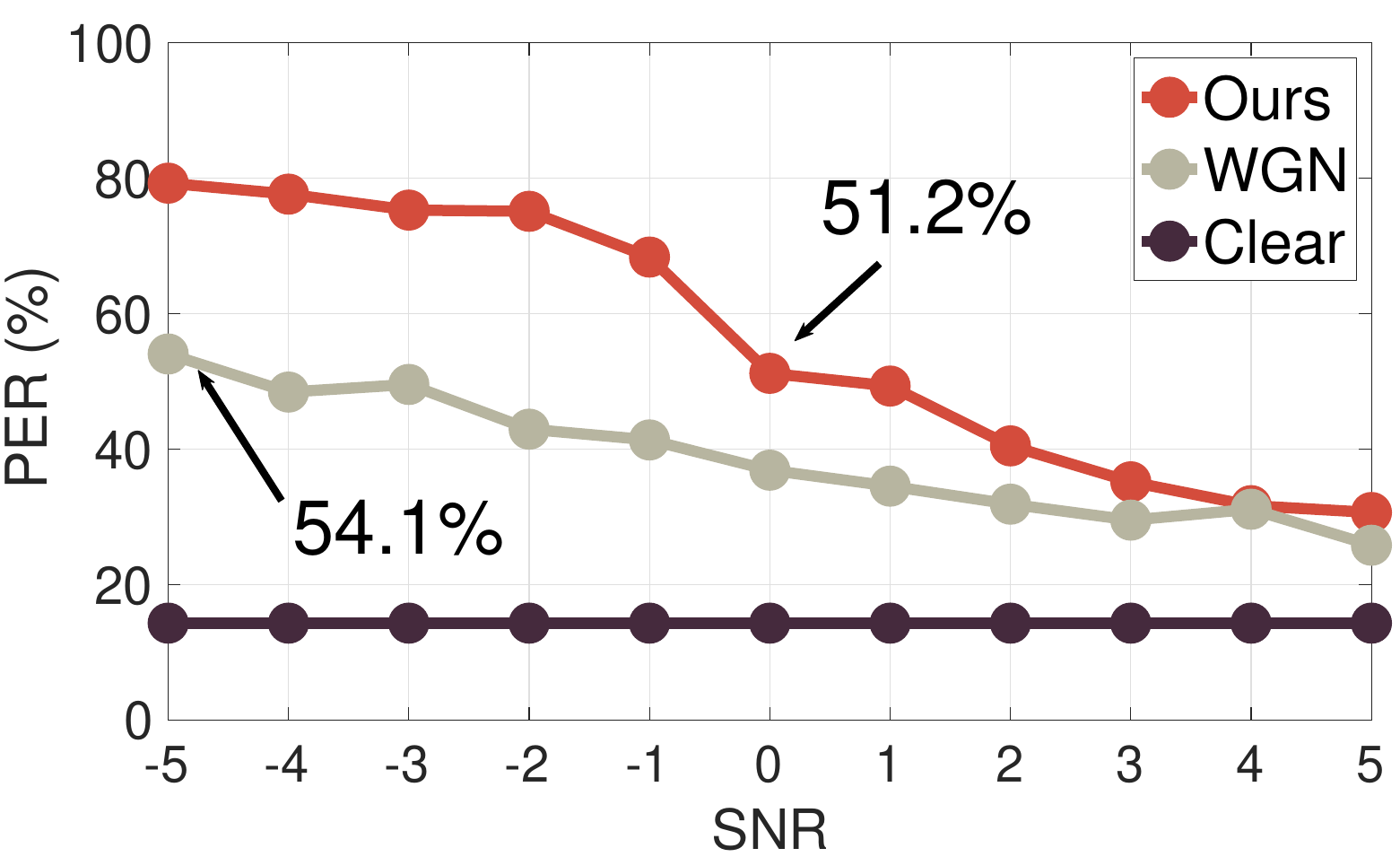}
        \caption{PER of specialized ASR systems.}	
        \label{fig:train_targeted_ASR}
    \end{minipage}
    \vspace{-5mm}
\end{figure}

\textbf{Normal Attacker with SOTA Speech Enhancement.} We first consider the attacker who has no knowledge of our system and could only use existing enhancement methods to improve the intelligibility of audio recordings. Here we consider a SOTA speech enhancement algorithm~\cite{hao2020fullsubnet} and enhance the speech signal obscured by different types of noise before recognition. We first employ WeNet for recognition and observe a notable improvement in accuracy for the WGN jamming case, surpassing $30\%$ in average, as shown in Figure~\ref{fig:digital_enhanced}. While for our noise, the accuracy decreases significantly after enhancement. We then further employ Tencent ASR and results show that the accuracy drops substantially after enhancement for both our noise and WGN. We speculate the decrease is caused by the conflict between Tencent ASR inherent speech enhancement and~\cite{hao2020fullsubnet}.

We further evaluate our noise robustness with the data collected in real-world scenario. The same enhancement method as before is used here and the result is shown in Tab~\ref{tab:real_world_enhanced}. Similar to the results in digital domain, the enhancement process reduces the recognition accuracy at each SNR.

\begin{table}[h]
    \caption{Robustness in real-world scenario against speech enhancement method}
    \centering
    \resizebox{\columnwidth}{!}{
        \begin{tabular}{cccccccc}
        \toprule
        \textbf{SNR}         & \textbf{\textless{}-4} & \textbf{{[}-4,-2{]}} &\textbf{ {[}-2, 0{]} }&\textbf{ {[}0,2{]}} &\textbf{{[}2,4{]}} & \textbf{\textgreater{}4 }\\ 
        \midrule
        \begin{tabular}[c]{@{}c@{}}Avg WER(\%)\\ (with Enhancement)\end{tabular} & 87.1          & 87.6        & 82.9        & 79.1      & 65.3      & 53.7\\ 
        \begin{tabular}[c]{@{}c@{}}Avg WER(\%)\\ (without Enhancement)\end{tabular} & 85.8          & 81.6        & 77.6        & 70.2      & 56.4      & 42.3\\ 
        \bottomrule
        \end{tabular}
    }
    \label{tab:real_world_enhanced}
\end{table}

\textbf{Advanced Attacker with Customized Speech Enhancement.} We then consider an advanced attacker that knows the detail of our jamming method. Instead of adopting existing enhancement methods, this attacker can train a model targeting on enhancing recordings jammed by our noise. Specifically, we consider the scenario described in Subsection~\ref{subsec:effectiveness_of_content_recovery}. The attacker creates a large speech dataset based on our jamming method and then trains the network as illustrated in Figure~\ref{fig:network_architecture} (without noise ref). Results in Figure~\ref{fig:content_recovery} (attacker part) shows that both the recognition accuracy and the audio quality (SI-SNR) decrease after speech enhancement, which means the advanced attacker can not recover the content from noisy recordings.

\textbf{Advanced Attacker with Specialized ASR.}
\label{subsec:robustness_adversarial_training}
We then consider an advanced attacker who targets on training a specialized ASR to recognize the noisy recording directly instead of enhancing it. Similar as before, We assume the attacker can create a large-scale dataset with our jamming method and train an ASR system from scratch. We choose TIMIT~\cite{timit} as the training data and take CRDNN~\cite{speechbrain} as the network architecture for its SOTA performance in speech recognition on TIMIT. The metric used here is PER (Phoneme Error Rate, similar to WER). We compare the recognition accuracy of specialized ASRs trained to recognize our noise and white noise respectively. Specifically, we train multiple ASRs and each one takes jammed speech signals with a specific SNR as the training data. As last, we have 22 ASRs considering the combinations of two types of noise and 11 SNRs.

The results are shown in Figure \ref{fig:train_targeted_ASR}. For white noise, the PER rises slowly with the decrease of SNR, which indicates the network's denoising ability for white noise. For our noise, the PER is slightly higher than white noise when $SNR\geq 0$, but increases rapidly when $SNR < 0$. These results mean that even when the attacker can train a specialized ASR, it is hard to recognize the audio jammed by our noise when $SNR \leq 1$, which is a reasonable value in real-world scenarios. We also find that when $SNR<0$, the training for the ASR targeting our noise can not converge properly for many reasons (e.g., exploding gradients), and requires careful parameter tuning.

\subsection{Comparisons with the Speech Noise}
\label{subsec:comparison_with_speech_noise}
\begin{figure}[t]
    \centering
    \includegraphics[width=\linewidth]{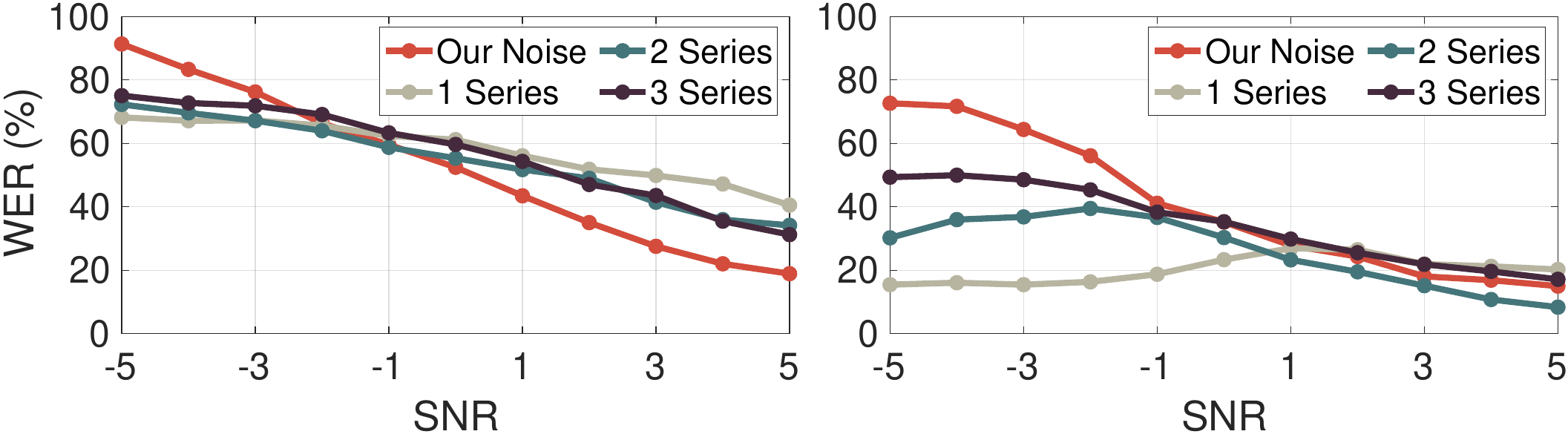}
    \caption{Comparisons with speech-like noise. Left: WER before noise reduction. Right: WER after noise reduction.}
    \label{fig:test_sep}
    \vspace{-3mm}
\end{figure}

In this part, we compare our noise with speech noises that contain 1, 2, and 3 speech series respectively. For a fair comparison, the speech series making up the noise is from the same people as the target. The results on the left of Figure~\ref{fig:test_sep} show that without considering noise reduction methods, our noise performs better when $SNR<-2$.

We then test the robustness of the noises against speech separation algorithms~\cite{sepformer} and trained three models targeting on noises containing 1, 2, and 3 series respectively. During the experiment, we find that although the separation process improves the intelligibility of audio for the human ear, the recognition accuracy of ASR is even worse than before. We think this is caused by the noise residue. So we further process the separated results with speech enhancement methods~\cite{fullsubnetPlus}. For our noise, we try to separate it with all three models separately then apply the enhancement. The result with the lowest WER is chosen for comparison. 

The result on the right of Figure~\ref{fig:test_sep} shows that when $SNR \ge 0$, the WERs of different tested noises are close; when $SNR<-1$, the WERs of speech noises with all SNRs are lower than 50\% and our noise performs much better. The result also reveals a phenomenon that higher noise energy for the speech noise does not always mean higher WER, which is different from our noise. Instead, the jamming performance of the speech noise will decrease when its noise energy is above certain threshold. The corresponding SNR thresholds are about 1, -3, and -4.8 for the noise containing 1, 2, and 3 speech series respectively. This phenomenon indicates that compared to our noise, the speech noise is less applicable in the real-world scenario as it is hard to control the noise energy to stay within a specific range.

\subsection{Comparisons with Other Jamming Methods}
Here we compare our noise with the counterparts in two other related works, namely Backdoor~\cite{backdoor} and Patronus~\cite{patronus}, and a commercial off-the-shelf device~\cite{commercial_device}. The noise in~\cite{patronus} consists of dynamic frequencies and chirps. The range of frequency is $[50, 40k]$ Hz and the duration of signal in each frequency is 0.2s. The noise in~\cite{backdoor} is a band-limited (i.e., $[0, 12k]$) white noise modulated by a $40$ kHz carrier. As for the commercial device~\cite{commercial_device}, we do not have the knowledge of its internals, therefore we can only speculate that its noise is made of variable multi-frequency tones based on the spectrogram. We conduct this experiment in the real-world scenario. We play audios with a smartphone and play noises with our transmitter (except for~\cite{commercial_device}) with constant power in the meantime. Then we record the audio with one smartphone placed at different distances from the transmitter. The recordings are enhanced with different methods before being fed into an ASR for better recognition (the same process as~\ref{subsec:comparison_with_speech_noise} for our noise and FullSubNet+~\cite{fullsubnetPlus} for others). The results in Figure~\ref{fig:comparison_with_other_noise} show that our noise performs better than others significantly.

\subsection{Impact of Data Augmentation Process}
\begin{figure}[t]
    \centering
    \begin{minipage}[t]{0.49\linewidth}
        \includegraphics[width = \linewidth]{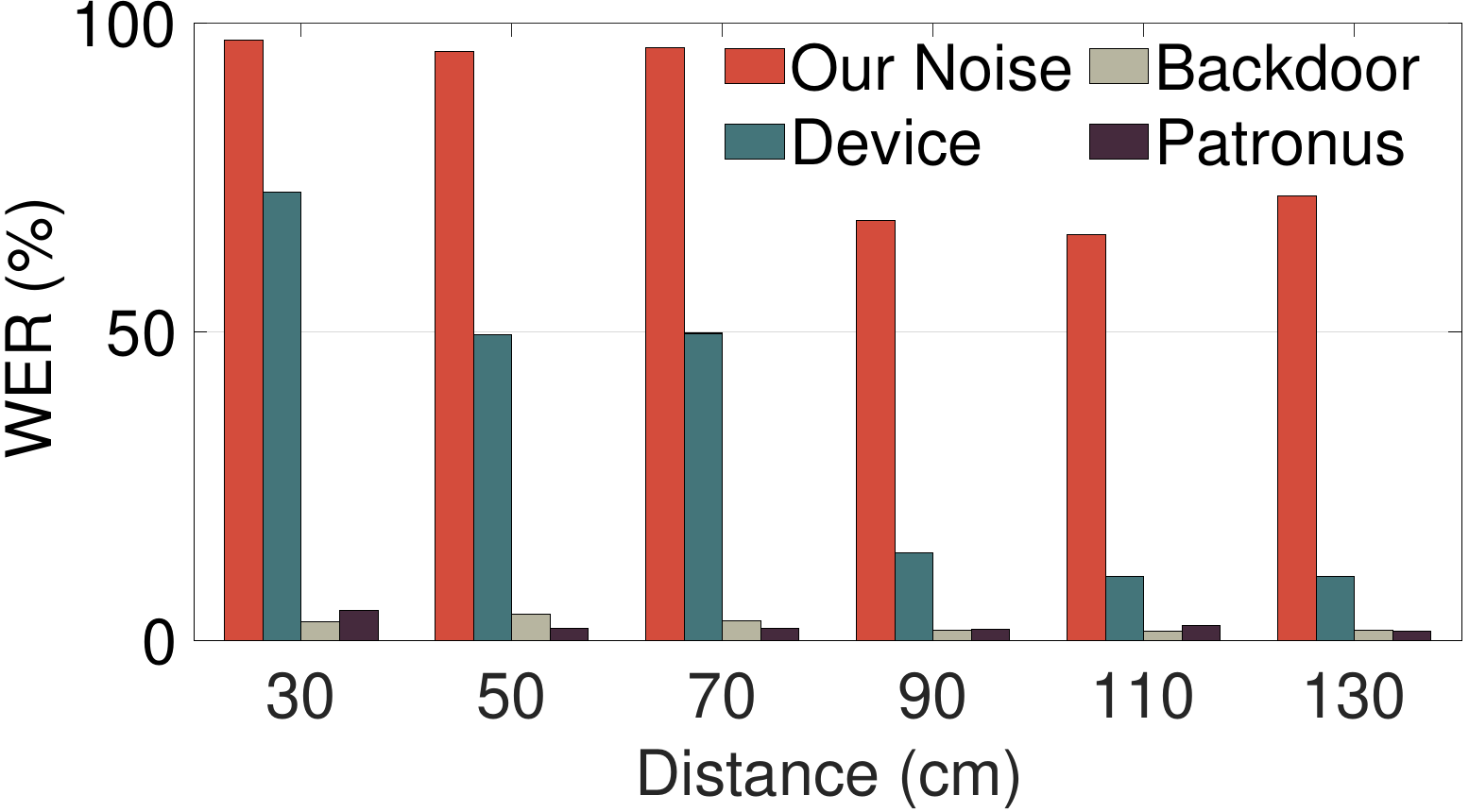}
        \caption{Comparisons with other existing methods}
        \label{fig:comparison_with_other_noise}
    \end{minipage}
   \begin{minipage}[t]{0.49\linewidth}
        \includegraphics[width = \linewidth]{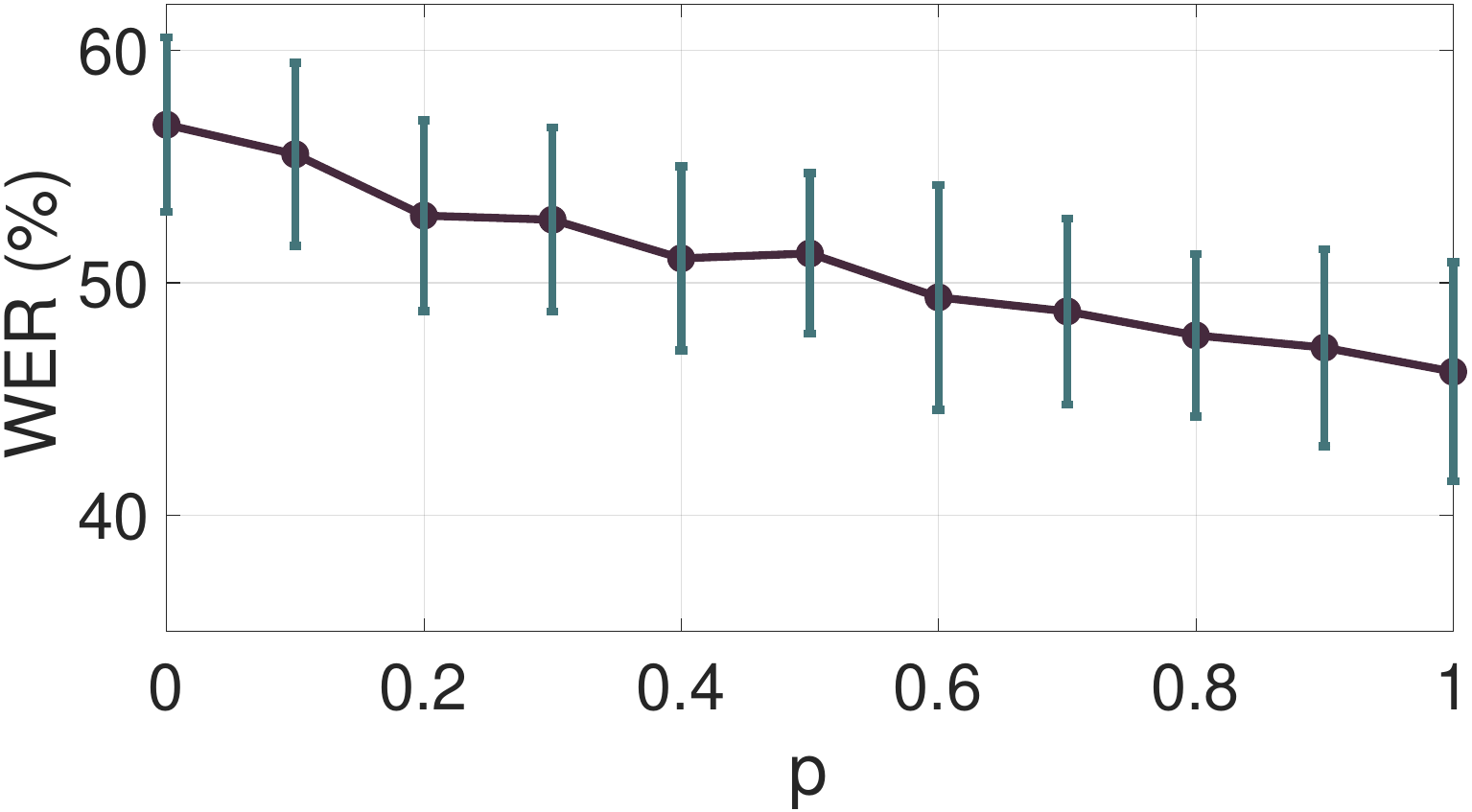}
        \caption{Impact of data augmentation}
        \label{fig:impact_of_augmentation}
    \end{minipage}
    \vspace{-5mm}
\end{figure}

Here we evaluate the impact of the data augmentation process on the effectiveness of our noise. In the noise generation process, before concatenating a newly selected phoneme data in to the noise, for each augmentation method, the phoneme data would be augmented with a probability $p$.
Then we test the jamming effectiveness of the noise generated with different $p$. The results in Figure~\ref{fig:impact_of_augmentation} show that as $p$ increases, the jamming effectiveness of our noise gradually decreases. However, the impact is limited, with only a drop of 5\% in WER when $p$ increases form 0 to 0.5.

\subsection{Impact of Recording Devices}
\label{subsec:impact_of_recording_devices}
\begin{figure}[t]
    \centering
    \includegraphics[width=\columnwidth]{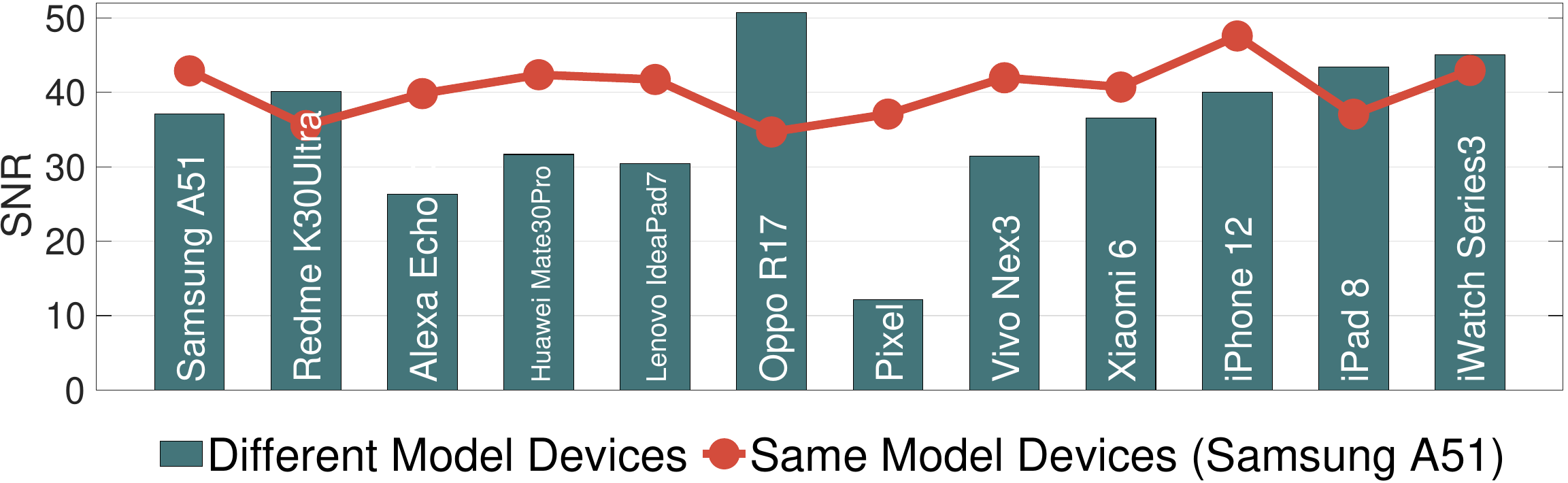}
    \caption{Nonlinearity in different recording devices}
    \label{fig:nonlinearity_in_diff_devices}
    \vspace{-3mm}
\end{figure}

To test the generalizability of InfoMasker among different recording devices, we use different appliances to record demodulated signals and them calculate their energy. We test a variety of recording devices, including nine models of smartphones, an iPad, a smart watch, a smart home device and a laptop. Besides, we also test the nonlinearity of twelve smartphones of the same model (Samsung A51). The results in Figure \ref{fig:nonlinearity_in_diff_devices} shows the good generalizability of InfoMasker.

\subsection{Case Study: A Common Office}
\label{subsec:case_study}
\begin{figure}[t]
    \centering
    \begin{subfigure}[b]{0.39 \columnwidth}{
        \centering
        \includegraphics[width=\linewidth]{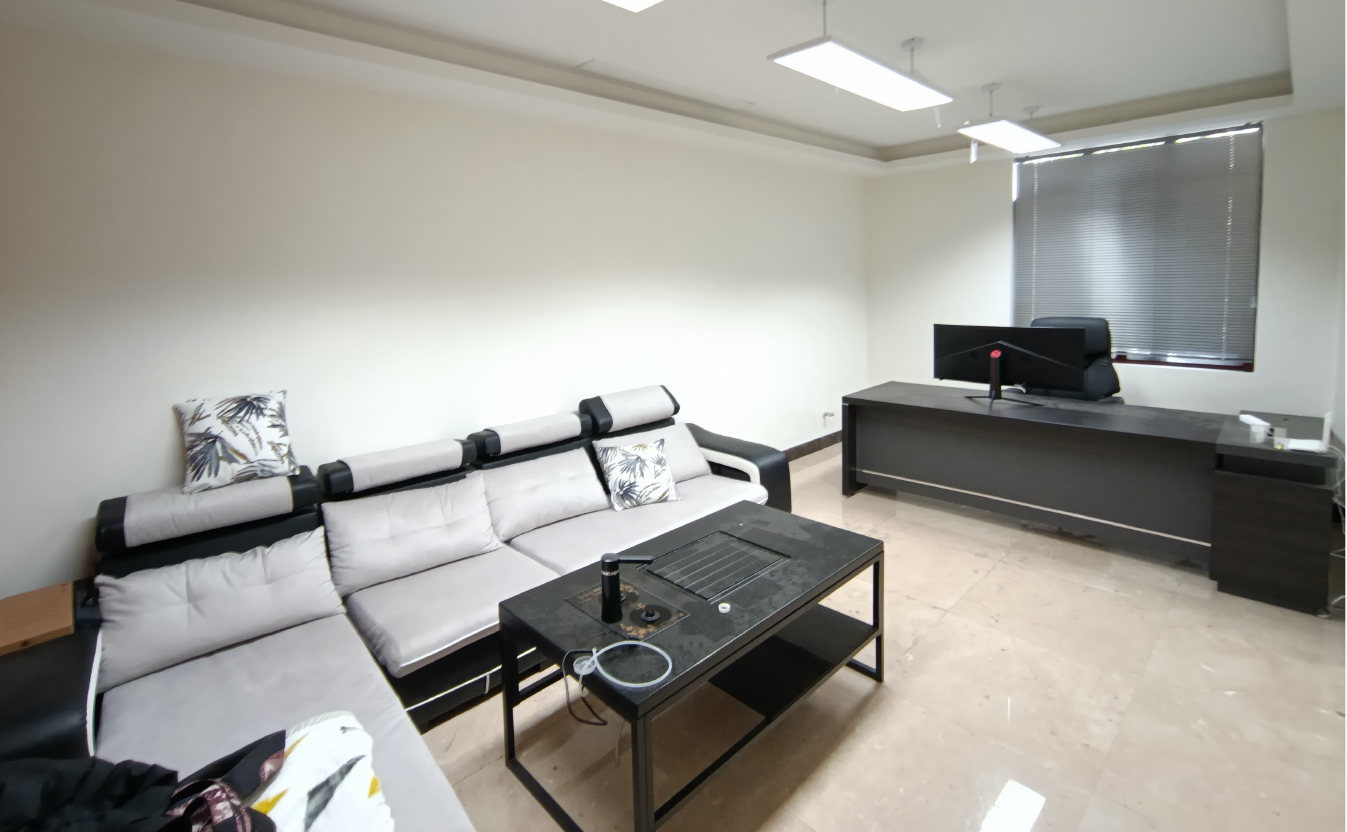}
        \caption{The office.}
        \label{fig:environment}
    }
    \end{subfigure}
    \begin{subfigure}[b]{0.59 \columnwidth}{
        \centering
        \includegraphics[width=0.8\columnwidth]{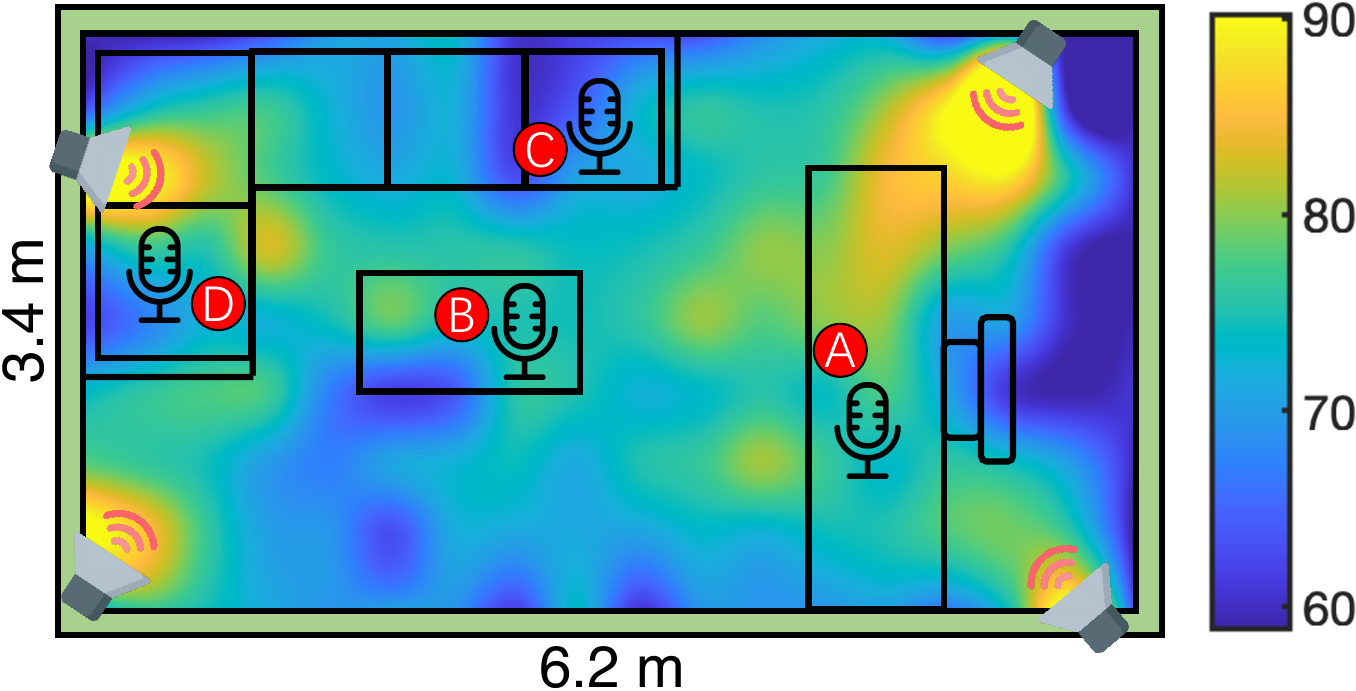}
        \caption{Energy Distribution.}
        \label{fig:room_distribution}
    }
    \end{subfigure}
    \caption{Experiment Settings.}
    \label{fig:equipment_environment}
    \vspace{-5mm}
\end{figure}

\textbf{Environment Setting.}
We further explore the possibility of deploying our system in a real-world environment. We deploy our system in a common office room, which is 6.2 meters long and 3.4 meters wide, as shown in Figure \ref{fig:environment}. We place a transmitter array in each of the four corners of the room. We first measure the ultrasound energy distribution in the room, and the result is shown in Figure \ref{fig:room_distribution}. 

Except for some areas within 10cm from the transmitter array, where the ultrasound energy reaches about 105 dB SPL, the energy in all other areas is lower than 95 dB SPL. This energy distribution meets the suggestion proposed by World Health Organization (WHO) that when humans are exposed to 40kHz ultrasound for more than 4 hours, the energy of the ultrasound should not exceed 110 dB SPL \cite{human_ultrasound_tolerance}.

\begin{figure}[t]
    \centering
    \begin{minipage}[t]{0.8\linewidth}
        \includegraphics[width = \linewidth]{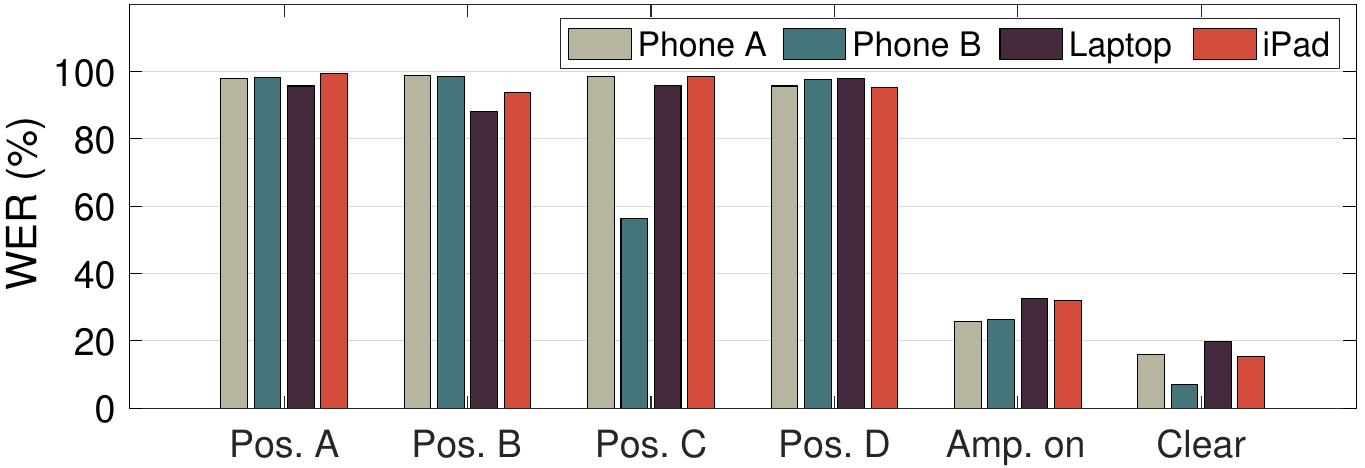}
    \end{minipage}

   \begin{minipage}[t]{0.8\linewidth}
        \includegraphics[width = \linewidth]{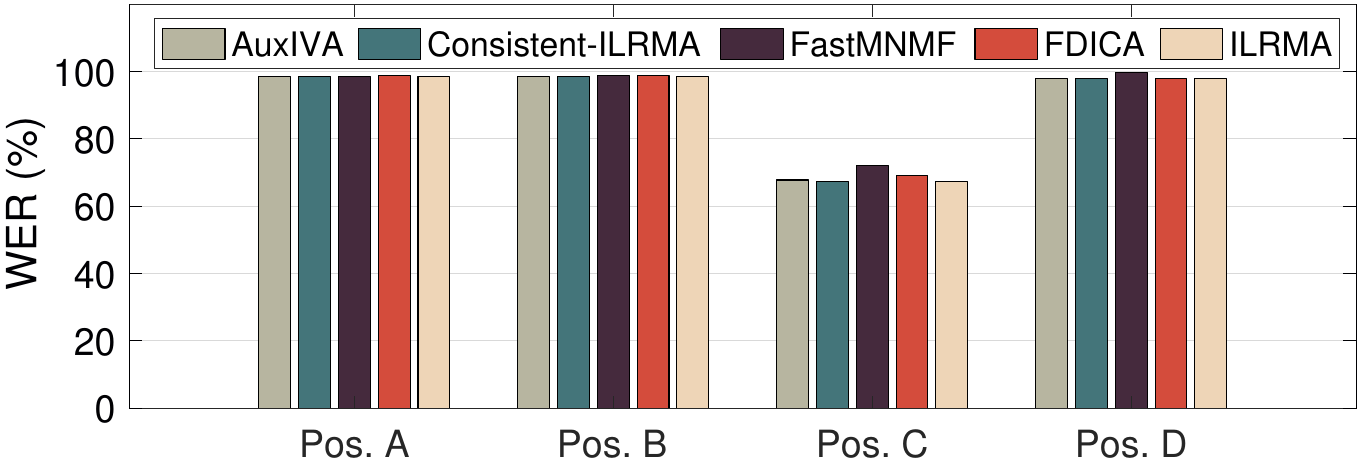}
    \end{minipage}

    \caption{Top: Recognition results for the case study. Bottom: Recognition results after BSS (Phone B).}
    \label{fig:case_study}
    \vspace{-5mm}
\end{figure}

\textbf{Recognition Results.}
We place four devices in the positions highlighted in red in Figure \ref{fig:room_distribution}. The noise energy in Point A and B is relatively high and is relatively low in Point C and D. The devices placed include two smartphones (Samsung A51 and Huawei Mate30Pro), a laptop (Lenovo IdeaPad7), and an iPad (iPad8). For each device we record about 3 hours data and the result is shown in Figure~\ref{fig:case_study}(top). We use three commercial ASR systems to recognize each data and choose the lowest WER as the result. For the reason that the power amplifiers will generate audible noise when turned on, we also test the scenario where the power amplifiers are turned on but no noise is transmitted.

\textbf{Blind Signal Separation.}
Considering the recording device may have multiple microphones or there are multiple devices recording at the same time, the attacker could use BSS methods to denoise the recordings. In the case study, the recordings from the Huawei Mate30Pro have two channels, so we test the effectiveness of BSS on these recordings. We test five BSS algorithms: AuxIVA \cite{AuxIVA}, ConsistentILRMA \cite{consistentILRMA}, FastMNMF \cite{fastMNMF}, LaplaceFDICA \cite{fdica}, and t-ILRMA \cite{ilrma}. The results are shown in Table~\ref{fig:case_study}(bottom). It can be noticed that even at position C, where the noise energy is the lowest, BSS still cannot improve the recognition accuracy.
\section{Related work}
\label{sec:related_work}

\textbf{Preventing Eavesdropping with Microphone's Nonlinearity. }
Several existing works attempt to jam microphones using ultrasound. Roy~et al. inject white noise to microphones using ultrasound \cite{backdoor}. Li~et al. generate noise with variable frequency according to a preset key to prevent the unauthorized devices from recording audios, while enabling the authorized users to recover speech signals from the noisy recordings \cite{patronus}. Sun~et al. propose MicShield, which can prevent the always-on microphones in smart home devices from recording private speech while passing the preset voice commands \cite{sun_alexa_2020}. Chen~et al. integrate ultrasonic transmitters into a wearable bracelet to expand the effective jamming coverage \cite{CHI}. However, the noises used in these works are either single tones with variable frequency or white noise, which are not robust when facing speech enhancement methods as demonstrated by the experiments in this paper.

\textbf{Other Applications with Microphone's Nonlinearity. }
Many works explore the microphone nonlinearity for other purposes, such as inaudible voice commands injection \cite{10.1145/3133956.3134052, 211283, yan_surfingattack_2020}, defense of inaudible commands \cite{he_canceling_2019, Zhang2021EarArrayDA, 211283}, communication \cite{backdoor}, and authentication \cite{zhou_nauth_2019}. Yan~et al. transmit the inaudible voice commands through solid medium to improve the stealthiness of the attack \cite{yan_surfingattack_2020}. Roy~et al. expand the attack range by striping different frequency bands to different transmitters \cite{211283}. Zhang~et al. exploit the difference of propagation attenuation between ultrasound and voice to distinguish inaudible commands injection from normal voice commands. Zhou~et al. validate that the parameters of the nonlinearity model in different microphones can be used as features for device authentication \cite{zhou_nauth_2019}.
\section{Conclusion}
\label{sec:conclusion}

In this paper, we propose InfoMasker, a highly effective anti-jamming system, which can prevent audio eavesdropping while preserving controlled recording privilege. By exploring informational masking effect, we achieve phoneme-based jamming noise design for the first time. Our noise exhibits strong ability to interfere with both ASR systems and the human auditory system, and it shows robustness against speech enhancement techniques. Moreover, our system optimizes the conventional ultrasonic transmission by using LSB-AM and integrating pre-compensation. Experiments show that our system can significantly obstruct the recognition accuracy of SOTA ASRs to below 50\% (SNR=0) and can provide recording privilege for authorized users. Our case study further validates the effectiveness of our noise in real-world scenario.  

\bibliographystyle{IEEEtran}
\bibliography{bib}

\end{document}